\documentclass[lettersize,journal]{IEEEtran}
\usepackage{enumerate,graphicx,epstopdf}
\usepackage{amsmath,amssymb,bm,amsthm}
\usepackage{cite}
\usepackage{mathtools}
\usepackage{array}
\usepackage{booktabs}
\usepackage[hyphens]{url}
\usepackage[dvipsnames, table]{xcolor}
\usepackage{flushend}

\usepackage[hidelinks,unicode]{hyperref}
\usepackage[english]{babel} 
\usepackage{makecell}

\usepackage[ruled,vlined]{algorithm2e}

\usepackage[overload]{empheq}
\usepackage{subcaption}
\usepackage{multirow}
\let\labelindent\relax
\usepackage{enumitem} 
\newcommand\scalemath[2]{\scalebox{#1}{\mbox{\ensuremath{\displaystyle #2}}}}

\newtheorem{rmk}{Remark}

\newtheorem{dfn}{Definition}

\newtheorem{prp}{Proposition}

\newcommand{\R}{\mathbb{R}}

\newcommand{\mc}[1]{\mathcal{#1}}
\newcommand{\bb}[1]{\mathbb{#1}}

\DeclareMathOperator*{\argmax}{arg\,max}
\DeclareMathAlphabet{\mymathbb}{U}{BOONDOX-ds}{m}{n}

\graphicspath{{figs/}}

\newcommand{\sh}[1]{\textcolor{RedViolet}{[#1]\raise 0.5ex \hbox{\footnotesize{SH}}}}
\newcommand{\shA}[1]{\textcolor{NavyBlue}{[#1]\raise 0.5ex \hbox{\footnotesize{SH}}}}

\newcommand{\fm}[1]{\textcolor{RedViolet}{[#1]\raise 0.5ex \hbox{\footnotesize{FM}}}}
\newcommand{\gb}[1]{\textcolor{OliveGreen}{[#1]\raise 0.5ex \hbox{\footnotesize{GB}}}}

\usepackage{caption}
\begin{document}

\title{Carbon-Aware Computing for Data Centers\\ with Probabilistic Performance Guarantees}
\author{Sophie Hall, Francesco Micheli, Giuseppe Belgioioso, Ana Radovanovi\'{c}, Florian D\"orfler 
\thanks{S. Hall, F. Micheli and F. D\"orfler are with the Automatic Control Laboratory, ETH Z\"{u}rich, 8092 Z\"{u}rich, Switzerland (e-mail: \texttt{\{shall, frmicheli, doerfler\}@ethz.ch}), G. Belgioioso is with KTH Royal Institute of Technology, 11428 Stockholm, Sweden (e-mail: \texttt{giubel@kth.se}), and A. Radovanovi\'{c} with Google Inc., Mountain View, CA 94043 USA (e-mail: \texttt{\{anaradovanovic\}@google.com}). This work was supported by the NCCR Automation, a National Centre of Competence in Research, funded by the Swiss National Science Foundation (grant number 51NF40\_225155), and by the Wallenberg AI, Autonomous Systems and Software Program (WASP) funded by the Knut and Alice Wallenberg Foundation.}}

\markboth{Journal of \LaTeX\ Class Files,~Vol.~14, No.~8, August~2021}%
{Shell \MakeLowercase{\textit{et al.}}: A Sample Article Using IEEEtran.cls for IEEE Journals}


\maketitle

\begin{abstract}
Data centers are significant contributors to carbon emissions and can strain power systems due to their high electricity consumption. To mitigate this impact and to participate in demand response programs, cloud computing companies strive to balance and optimize operations across their global fleets by making strategic decisions about when and where to place compute jobs for execution.
In this paper, we introduce a load shaping scheme which reacts to time-varying grid signals by leveraging both temporal and spatial flexibility of compute jobs to provide risk-aware management guidelines and job placement with provable performance guarantees based on distributionally robust optimization.
Our approach divides the problem into two key components: (i) day-ahead planning, which generates an optimal scheduling strategy based on historical load data, and (ii) real-time job placement and (time) scheduling, which dynamically tracks the optimal strategy generated in (i).
We validate our method in simulation using normalized load profiles from randomly selected Google clusters, incorporating time-varying grid signals. We can demonstrate significant reductions in carbon cost and peak power with our approach compared to myopic greedy policies, while maintaining computational efficiency and abiding to system and grid constraints.

\end{abstract}


\section{Introduction}

The number of hyperscale data centers (DCs) doubled between 2015 and 2021~\cite{masanet2020recalibrating}. In 2022, energy demand from DCs and data transmission networks accounted for about 1.5\% of the global electricity demand (240-340 TWh) and for 1\% of the energy-related greenhouse gas emissions~\cite{iea2023data}. Despite the rapid growth in the number of DCs, their global power demand and related greenhouse gas emissions have remained nearly constant between 2015 and 2020, largely due to significant improvements in operational efficiency of individual data centers~\cite{gsg2024generational}. However, efficiency gains are stagnating, and with the rising demand of AI technologies, ever more capacity is needed. As a result, global electricity demand from data centers is expected to triple between 2020 and 2030~\cite{gsg2024generational}, placing substantial strain on local grid infrastructure~\cite{hodgson2024booming}. Nevertheless, well-coordinated networks of DCs can also provide flexibility to the power system by participating in demand response (DR) programs and offering ancillary services~\cite{enelx2024how}.

Cloud computing companies usually own large data center fleets spread across the globe. For example, in 2021, Microsoft, Amazon, and Google collectively owned more than 50\% of all hyperscale DCs~\cite{srg2021microsoft}. As efficiency gains at individual DCs begin to plateau and flexibility provision to the grid is rewarded, these companies are looking for ways to jointly optimize operations across their entire fleet to reduce operational costs and carbon footprint to reach their sustainability goals~\cite{gsg2024generational, johnson2020carbon, ramachandran2024announcing,dixit2024retinas,verrus2024how}, such as Google's goal of net-zero emissions by 2030~\cite{google2024net}. Moreover, optimizing their global fleet will allow cloud computing companies to participate in DR schemes, unlocking new revenue streams and supporting the power system as a result. Several studies in Europe~\cite{bovera2018economic, hansson2022potential} and the U.S.~\cite{wierman2014opportunities} have explored the potential for grid stabilization through their participation in DR programs. Google has recently performed DR pilot studies across Asia, Europe and the U.S.~\cite{mehra2024using}.

It is well-known that certain types of compute jobs, such as offline data processing, model training and simulation pipelines, are temporally flexible, while they consume significant computing resources. At the same time, large global balancing systems serving user requests have the potential to become carbon-aware and grid-aware through smart coordination.
A key enabler for cloud computing companies to optimize operations and respond to time-varying grid signals is to exploit the spatial and temporal flexibility of different classes of compute jobs - meaning that they can be executed in different DC locations and/or delayed~\cite{xu2020managing,abusharkh2017optimal, dvorkin2024agent}. 

Whether reacting to DR signals or internally optimizing a global DC fleet, its time-varying operation is a challenging problem for multiple reasons: (i)~it requires thousands of placement decisions every second, with compute jobs arriving continuously throughout the day; (ii)~data on individual compute jobs is not accessible before submission; and (iii)~job placement within clusters (with each data center containing multiple clusters) is managed by low-level operating systems, which are difficult to model and modify by high-level planners.

The literature covers various aspects of the problem but has several key limitations. Specifically, most works (i) do not exploit temporal and spatial flexibility jointly~\cite{chen2017dglb,liu2015greening, lindberg2022using, paul2013price}; (ii) rely on either optimization-based planning approaches, which are not tractable in real-time operation~\cite{lindberg2022using, breukelman2024carbon}, or myopic greedy policies, which are suboptimal as they do not take into account future grid signals or compute load predictions~\cite{wang2018optimal,xu2020managing, khosravi2017dynamic}; and (iii) either assume exact knowledge of incoming compute jobs, which is unrealistic, or lack specifics on real-time job placement~\cite{lindberg2022using, dvorkin2024agent}.

The authors in~\cite{lindberg2022using} propose geographic load shifting by solving a bilevel program based on locational marginal carbon intensities. While promising, marginal carbon intensity data is not readily available in practice and it is unclear how the scheme would be applied in real-time operation. In~\cite{dvorkin2024agent}, a bilevel program combines power system planning with a latency-minimizing task shifting problem, approximating the solution through linear regression. However, for a practical implementation, considering the stochasticity of compute load and giving robust service guarantees, would be essential.

Most real-time job-placement schemes rely on lightweight but inherently myopic greedy rules.  For example, \cite{nadalizadeh2021greenpacker} chooses, at every job arrival, the cluster that minimizes a weighted score of instantaneous renewable supply, electricity price, power-of-usage effectiveness, and server fragmentation, achieving improved energy-cost savings without any look-ahead. The work in~\cite{goiri2015matching} maximizes on-site solar utilization by deferring deadline-flexible batch jobs to daylight hours.  
The authors in~\cite{aksanli2016renewable} push this idea further: they combine dynamic electricity prices with minute-ahead PV ``now‑casting'' and route workloads among geographically separated data centers that are expected to have the cheapest (or greenest) power supply in the near future.  
The authors in~\cite{khalil2019energy} apply an $O(1)$ rule that directs each request to the site minimizing an energy-delay product, thereby trading off incremental energy use against network-induced latency.  
Although all four heuristics are computationally cheap, they optimize each placement in isolation and provide no formal guarantees on efficiency or robustness under stochastic compute-load conditions.


In this work, we build on the model recently presented by Google~\cite{radovanovic2023carbon}, which leverages temporal flexibility of compute jobs to minimize the expected carbon footprint and daily peak power over the entire fleet of DCs. In this model, temporal shifting of compute jobs is indirectly regulated by introducing \textit{Virtual Capacity Curves} (VCCs), which are hourly limits that artificially cap the computing resources available to the real-time job placement algorithm in each cluster.



If cloud computing companies want to participate in DR schemes, they will have to redistribute compute jobs while maintaining a guaranteed high level of service. However, as compute load of incoming jobs is highly stochastic, any redistribution and scheduling scheme must ensure robustness with respect to job queuing latency. Distributionally robust optimization (DRO) is an optimization technique which inherently offers robustness against unexpected compute load profiles, rare events, and distribution shifts that might be caused by changes in the job submission pattern.





We propose a distributionally-robust day-ahead planning problem coupled with a real-time placement algorithm which is driven by data, offers robustness guarantees, and considers various degrees of flexibility of jobs while explicitly incorporating grid codes and demand response (DR) events. Integrating stochasticity of DR events is very relevant as Google has been called in the past to drop load in emergency situations~\cite{mehra2024using}. To summarize, our approach is unique in the following ways:
\begin{enumerate}
    \item Temporal and spatial flexibility: Our approach accommodates jobs with any flexibility, from 0 to 24h delay tolerance and local to global execution range, whereas in~\cite{radovanovic2023carbon} only the temporal flexibility was considered.
    \item Data-driven: We directly approximate the probability distribution of compute loads from data within the planning problem, without making assumptions on the distribution. 
    \item Tuning robustness: The probabilistic performance guarantees are tunable within the distributionally-robust optimization problem allowing to trade-off reliability and performance. All types of jobs are treated equally and performance guarantees hold across compute job classes.
    \item Co-design of VCCs and the scheduling strategy: Deciding when and where jobs are being sent while limiting cluster capacity allows to fully exploit the spatial and temporal correlations in the data and perform preemptive peak planning such that spatially flexible loads go to underused clusters, thus reducing job queuing latency.
    \item Grid requirements: We incorporate grid codes and demand response services in the modeling of the scheduling problem in a probabilistic sense, to account for the unknown distribution of demand response events and compute loads.
\end{enumerate}

The proposed approach provides rigorous guarantees based on DRO theory while abiding to the key constraints discussed in~\cite{radovanovic2023carbon} and from our discussions with Google. 
We validate our approach using a sample of normalized load profiles from randomly selected Google clusters and show that it outperforms commonly used greedy policies in terms of peak power and carbon cost while providing theoretical guarantees.




Although this work focuses on real-time operation of data centers, the ability to reliably quantify and manage flexible compute demand has direct implications also for long-term planning in power systems. For example, when incorporated into generation–transmission expansion models, the spatio-temporal flexibility of data centers can (i) cut peak demand and thus defer or reduce new generation capacity~\cite{TejadaArango2020TPWRS}; (ii) increase the use of spatially diverse wind and solar, informing renewable siting and curtailment mitigation~\cite{Asensio2018TSG}; and  
(iii) enable co‑optimized expansion models in which demand flexibility lowers total system costs~\cite{VazquezPombo2023TSG}.  
Hence, the operational approach developed here provides not only short-term carbon and cost benefits but also actionable inputs for the long-term integrated planning of grid infrastructure and compute facilities.

The rest of the paper is structured as follows. We introduce basic notation and state preliminaries in Section~\ref{sec:Preliminaries}. In Section~\ref{sec:ProblemSetting}, we introduce the model and formulate the stochastic optimization problem. In Section~\ref{sec:DROAllocation}, we derive a distributionally-robust load schedule, approximating the ambiguity set from historical data. Section~\ref{sec:RTPlacement} discusses real-time job placement. Section~\ref{sec:Numerics} and Section~\ref{sec:PracticalImp} present simulation results and Section~\ref{sec:Conclusion} concludes the paper. 


\section{Notation and Preliminaries}\label{sec:Preliminaries} 

    We denote the set of the first \( K \) positive integers as \( \bb{Z}_K := \{1, \dots, K\} \), and the set of positive integers from \( t \) to \( T \), with $t<T$, as \( \bb{Z}_{[t:T]} := \{t, \dots, T\} \). The operator \( [\,\cdot\,]_{+} := \max\{\cdot, 0\} \) is the projection on the positive orthant. Given some variables \( v_{t,d} \), with \( t \in \mc{T} \) and \( d \in \mc{D} \), we denote the stacked vector as \( v := \text{col}(v_{t,d})_{t \in \mc{T}, d \in \mc{D}} = [v_{1,1}, \dots, v_{T,D}]^\top \). Table~\ref{tb:Notation} summarizes all parameters, sets, and variables used.

\begin{table}[ht]\renewcommand{\arraystretch}{1.2} 
\caption{Modeling notation \label{tb:Notation}}
\begin{tabular}{@{}lp{4.0cm}@{}}
\toprule
Indices, Sets and Parameters \\ \midrule
$k \in \mc{H}:=\{1,\dots,K\}$ [hrs]& Submission time\\
$t \in \mc{T}:= \{1,\dots,T\} $ [hrs]& Execution time \\
$d\in \mc{D}:=\{1,\dots,D\}$&  Data center cluster index\\
$c \in \mc{C}:= \{1,\dots,C\}$&  Job class\\
$\mc{D}_{c}\subseteq \mc{D}$&  Set of clusters at which jobs of class $c$ can be allocated\\ 
$h_c$ [hrs]&  Time flexibility of class $c$ \\
$Y\in \R^{K\times C\times T\times D}_{\geq 0 }, Y_{k,c,t,d}\in \R_{\geq 0}$, & Scheduling strategy as tensor and individual entry\\
$y\in \R_{\geq 0}^{K C T D}$, & Scheduling strategy as vector\\
$v_{t,d}$ & Virtual Capacity Curve \\
$\overline{v}_{t,d}$ & Real (cluster) machine capacity\\
$p_{t,d}$ & Power load\\
$\bar{p}_{d}$ & Maximum power limit\\
$\bar{p}_{d}^{\uparrow/\downarrow}$ & Maximum ramp up/down power limit\\
$\Delta t$ & Time conversion factor from ramping to planning time scales\\
$\alpha_{1,d}$,  $\alpha_{0,d}$ & Slope and intercept of the affine power model\\
$\rho^{\text{carb}}_{t,d}$ & Carbon cost metric\\ 
$\rho_d^{\text{in}}$ & Infrastructure cost metric\\\bottomrule
\end{tabular}
\end{table}
In the following, we introduce relevant concepts from distributionally robust optimization (DRO)~\cite{kuhn2019wasserstein}. The conditional value at risk (CVaR) is a popular risk measure that provides the average of the tail end of the loss distribution. Intuitively, it captures the expected value of the worst-case $\beta$ fraction of outcomes with $\beta \in (0,1]$. It is defined as follows.
\begin{dfn}\label{dfn:CVAR}
For a random variable $\omega \in \Omega \subset \R^r$ with distribution $\bb{P}_{\omega}$ and a function
$\phi : \R^r \mapsto \R$, the CVaR of level $\beta$ is defined as
\begin{align}
CVaR_{1-\beta}^{\omega \sim \mathbb{P}_{\omega}}(\phi(\omega)):= \inf_{q\in \R}[\beta^{-1} \bb{E}^{\omega \sim \bb{P}_m} [[\phi(\omega) + q]_{+}] -q].
\end{align}
\end{dfn}
The Wasserstein metric quantifies the minimum cost required to transport one distribution $\bb{Q}_1$ into another $\bb{Q}_2$.
\begin{dfn}[Wasserstein distance~\cite{esfahani2017data}] \label{dfn:WasserMet}
 Consider distributions $\bb{Q}_1, \bb{Q}_2 \in \mc{M}(\mc{Y})$  where $\mc{M}(\mc{Y})$ is the set of all
probability distributions $\bb{Q}$ supported on $\mc{Y} \subseteq \bb{R}^{nT}$ such that $\bb{E}[ \|y\|]< \infty$. The Wasserstein metric $d_W : \mc{M}(\mc{Y}) \times
\mc{M}(\mc{Y}) \to R_{\geq 0}$ between the distributions $\bb{Q}_1$ and $\bb{Q}_2$ is defined as
\begin{equation}
d_W(\bb{Q}_1,\bb{Q}_2) := \inf \left\{ \int_{\mc{Y}^2} \/ \|\bf{y}_1 -\bf{y}_2 \| \Pi (\mathrm{d}\bf{y}_1, \mathrm{d}\bf{y}_2)\right\},
\end{equation}
where $\Pi$ is the joint distribution of $\bf{y}_1$ and $\bf{y}_2$ with marginals $\bb{Q}_1$ and $\bb{Q}_2$, respectively.
\end{dfn}
This distance metric $d_w$ is also referred to as the ``type-1 Wasserstein distance".
\section{Problem setting}\label{sec:ProblemSetting}
We consider the problem of dynamically allocating an aggregate compute load, denoted by $s$, among clusters $\mc{D}:=\{1,\dots,D\}$ geographically distributed across locations, or data centers. Each data center contains multiple clusters. The compute load comprises resource usage across individual compute jobs that get placed in clusters over a 24-hour (planning) horizon, at one-hour intervals $k$, with $k \in \mathcal H := \{1,\dots,K\}$, and leave the system upon job completion or cancellation. We consider compute jobs that tolerate delayed execution, e.g., data processing pipelines, log analysis, large-scale simulations, and nightly builds~\cite{dean2004mapreduce}. These jobs vary in terms of their temporal and spatial flexibility. To categorize them, we define a set $\mc{C}:= \{1,\dots,C\}$ of flexibility classes, wherein each class $c \in \mc C$ is associated with:
\begin{enumerate}
\item[(i)] a temporal flexibility $h_c \in \mathbb Z_{\geq0}$, namely, the maximum delay (in hours) the compute jobs can tolerate,
\item[(ii)] a spatial range, encoded via $\mc{D}_c\subseteq \mc{D}$ consisting of all clusters at which jobs of class $c$ are allocable\footnote{The spatial range is a function of the compute job recurrence, network consumption, data dependence and more.}.
\end{enumerate}
Crucially, the exact future aggregate load profile of each class $c$ is unknown at the beginning of the planning horizon. Hence, we treat it as a random vector $ s_c $ drawn from an underlying probability distribution, denoted by \(\mathbb{P}\). Each element \(s_{c,k} \), with $k \in \mc H$, within this random vector represents the aggregate load of compute jobs of class $c$ submitted at time $k$ for processing.
\begin{rmk}
    Classifying the spatial and temporal flexibility of compute jobs is challenging and an active area of research. Choosing the number of classes involves a trade-off between per-class prediction accuracy, computational complexity, and performance gains with increasing class granularity.
\end{rmk}
Our goal is to design a scheduling strategy that is  practical for real-world deployment, and leverages the temporal and spatial flexibility of compute jobs to ensure the DC network operates reliably (i.e., respecting capacity limits) and efficiently (i.e., quickly allocating jobs while minimizing carbon footprint), despite the uncertainty of next day's compute load.

\smallskip
We represent the scheduling strategy for these compute loads with a tensor $Y \in \mathbb{R}^{K \times C \times T \times D}$, where each entry $Y_{k,c,t,d} \geq 0$, describes the fraction of the aggregate load $s_{c,k}$ of class $c$, submitted at time \(k\), that is allocated for processing at time $t$ to cluster $d$. Notably, $t \in \mc{T}:= \{1, \dots, T\}$ with $T= K + \max_{c\in \mc C}h_c$ represents the planning horizon length for job execution. This horizon length must be larger than the submission horizon because jobs of class $c$ submitted at time $K$ still have a time flexibility of $h_c$. 


\subsection{Compute load constraints}
As the scheduling strategy assigns fractions of compute load to multiple clusters and times, it is essential to ensure that these fractions sum up to one. This guarantees that the entire load is accounted for, as captured by the following constraint:
\begin{align} 
\label{eq:TSconstr}
\sum_{d \in \mathcal{D}} \sum_{t\in\mathcal{T}} Y_{k,c,t,d}\, =\, 1, \quad \forall k \in \mathcal{H}, \; c \in \mathcal{C}.
\end{align}
Since each class $c$ has limited spatial range and temporal flexibility, the fraction of its load scheduled for clusters outside the spatial range, $d\in  \mc{D}\backslash \mc{D}_c$, and all hours before the submission time $k$ and after the delay tolerance $k+h_c$, must equal zero. This requirement is formalized as follows: 
%
\begin{subequations} \label{eq:SpatTempConst}
\begin{align} \label{eq:SpatConst}
    &Y_{k,c,t,d} = 0, \quad \forall k\in \mc{H}, c\in \mc{C}, t\in \mc{T}, d\in  \mc{D}\backslash \mc{D}_c,\\
    \label{eq:TempConst}
    &Y_{k,c,t,d} = 0, \quad  \forall k\in \mc{H}, c\in \mc{C}, t\in \mc{T} \backslash  \bb{Z}_{[k:k+h_c]},d\in  \mc{D}.
\end{align}    
\end{subequations}
%
%
\begin{figure}[t]
\centering 
\includegraphics[width=\columnwidth]{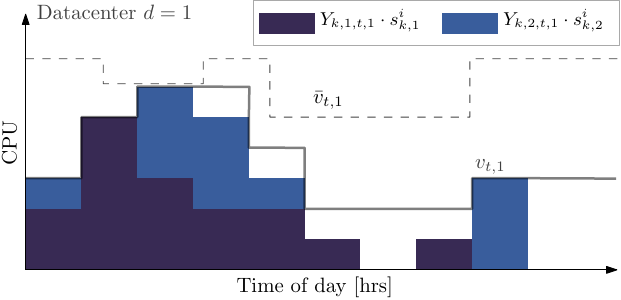}
\caption{An example of the aggregate load schedule for compute loads from two flexibility classes $c\in \{1,2\}$ at a single cluster $d=1$ over $24$ hours. The VCC (solid line) $v_{t,1}$ limits the allocable load at each time interval $t$, while the true capacity $\overline{v}_{t,d}$ (dashed line) is obtained by subtracting the inflexible load$^2$ from the (cluster) machine capacity.} \label{fig:VCC}
\end{figure}
%

%
The operational capacities $v_{t,d}$  of each cluster $d$, known as \textit{virtual capacity curve} (VCC), determine the largest load allocable at a cluster. The VCC limits induce indirect temporal shifting of compute jobs as low capacity during times of peak compute load will lead to jobs being queued and their execution being delayed.

Inherently, each $v_{t,d}$ is limited by the true capacity $\overline{v}_{t,d}$, which takes into account the (cluster) machine capacity and the predicted inflexible load\footnote{Inflexible load consists of compute jobs that cannot be shifted in time and across-locations.} yielding the following constraint
\begin{align}\label{eq:VCC_constraint_with_base_load}
0\leq v_{t,d} \leq \overline{v}_{t,d},  \quad \forall t\in \mc{T}, \, d \in \mc{D}.
\end{align} 

In the current Google model~\cite{radovanovic2023carbon}, the VCCs were computed independently of the job scheduling strategy. In this paper, we treat $v = \text{col}(v_{t,d})_{d\in\mc{D}, t\in\mc{T}}$ as an additional decision variable and propose the co-design of VCCs and scheduling strategy $Y$. 
Intuitively, the allocated aggregate load per cluster $d$ and time $t$ must be lower than the VCC limit $v$, thus we define the following constraint function 
\begin{align}\label{eq:g1}
&g_1(Y,s,v) = \max_{\substack{t\in \mc{T},\\d\in \mc{D}}}\left(
\sum_{c \in \mathcal{C}} \sum_{k \in \mathcal{H}} Y_{k,c,t,d} \cdot s_{k,c} - v_{t,d}\right)
\end{align}
Since the compute loads $s_{k,c}$ are random variables, we enforce this constraint in a probabilistic sense as following using the following Conditional Value-at-Risk (CVaR) constraint: 
\begin{align}
\label{eq:CConstrPr}
&\text{CVaR}_{1-\beta}^{s \sim \mathbb{P}}\left[\max_{\substack{t\in \mc{T},\\d\in \mc{D}}}g_1(Y,s,v)
\right] \leq  0.
\end{align}
Loosely speaking, this constraint ensures that, for each cluster $d$ and time step $t$, a large percentile (tuned by $\beta$) of random load profiles $s_{c,k}$ allocated using the strategy $Y_{k,c,t,d}$ satisfies the capacity constraint $v_{t,d}$. The CVaR operator in \eqref{eq:CConstrPr} is formally introduced in Definition~\ref{dfn:CVAR}.

\subsection{Power constraints}
o effectively manage power consumption of a cluster, it is necessary to model the relationship between compute load and power. We will make use of the power model that is running in production and relates CPU usage of Google’s Power Distribution Units (PDU) to their power consumption~\cite{radovanovic2022power}. A single PDU is comprised of thousands of machines, and a handful of PDUs comprise a cluster. Although DC infrastructure planning and real-time job placement depend on multi-dimensional resource usage (CPU, RAM, disk usage), it was demonstrated in~\cite{radovanovic2022power} that an affine model applies for PDUs with different machine types (e.g. standard compute, storage, networking). Therefore, we can write
\begin{align*}
    p^{\text{PDU}} = \alpha_1^r\, l^{\text{PDU}} + \alpha_0^r, 
\end{align*}
where $l^{\text{PDU}}$ is the CPU usage at a PDU and the slope $\alpha_1^r$ and intercept $\alpha_0^r$ would be fitted for each PDU and usage regimes. The idle power of machines depends on the platform family and lies between 30\%-40\% of their maximum power consumption, i.e., $\alpha_0\in [0.3,0.4] \cdot p_{\text{max}}^{\text{PDU}}$. The slope is primarily influenced by the CPU usage and the workload mix running in the PDU, and it is in most cases smaller than 10 (most commonly smaller than 1), thus $\alpha_1^r\in [1,10]$. However, in this work, we are interested in the power consumption of an entire cluster at time $t$, denoted as $p_{t,d}$, which is the aggregate power consumption of all PDUs in a cluster. In practice, DCs are connected to the power grid and must adhere to grid codes, such as contractual maximum power consumption limits $\bar{p}_d$, to avoid getting their supply throttled. This yields the constraint
\begin{align}\label{eq:ConstrMaxPow}
   0\leq  \text{PUE} \cdot\,  p_{t,d}\; \leq\; \bar{p}_d,  \quad \forall t\in \mc{T}, \, d \in \mc{D}.
\end{align} 
where PUE is the Power Usage Efficiency factor which considers power required for cooling of the data center and translates the aggregate power consumption of PDUs into total grid-level power demand. In practice, Google maintains a nearly constant PUE of 1.09 across all data centers and seasons~\cite{google2024efficiency}.
\begin{rmk}
Google data centers rely on local controllers for cooling~\cite{gao2014machine} which achieve constant PUE levels across seasons and locations~\cite{google2024efficiency}.
Therefore, it is not necessary to explicitly model cooling power in our load shifting mechanism and incorporating the proportional PUE gain into \eqref{eq:ConstrMaxPow} suffices.
\end{rmk}
Another relevant grid constraint for DCs are ramping limits which specify how fast load may be dropped or increased without destabilizing the local grid. For instance the Swiss code specifies a linear change over a period of 10 minutes~\cite{swissgrid2019transmission}. The exact time scales and limits specified in the code differ across grids. Yet, most of them can be modeled using a linear constraint as follows
\begin{align}\label{eq:RampLim}
   \Delta t \, \bar{p}_d^{\downarrow} \leq  (p_{t+1,d} - p_{t,d})\; \leq\; \Delta t \, \bar{p}_d^{\uparrow},  \quad \forall t\in \mc{T}, \, d \in \mc{D}, 
\end{align}
where $\bar{p}_d^{\uparrow/\downarrow} $ are ramping up and down limits and $\Delta t$ is a time conversion factor between ramping and planning time scales. This approach is consistent with ramping limits considered in the literature~\cite{dvorkin2024agent}\footnote{Note that multiple clusters comprise one data center, and thus depending on the infrastructure set-up, DC limits would have to be converted to per-cluster limits.}. Similarly to the VCC limits, the power consumed at a given cluster is determined by the random, unknown load profile $s_{k,c}$ and the scheduling strategy $Y$. To enforce this as a probabilistic constraint, we define the function
\begin{align}\label{eq:g2}
    g_2(Y,s,p) = \max_{\substack{t\in \mc{T},\\d\in \mc{D}}}\left(  \alpha_{1,d}
\sum_{c \in \mathcal{C}} \sum_{k \in \mathcal{H}} Y_{k,c,t,d} \cdot s_{k,c} + \alpha_{0,d} - p_{t,d}\right),
\end{align}
where $\alpha_{1,d}$ is the slope and $\alpha_{0,d}$ the idle power consumption of an entire cluster $d$. Having both constraint functions $g_1$ defined in \eqref{eq:g1} and $g_2$ in place enables us to combine them in one probabilistic constraint using a single CVaR measure as follows:
\begin{align}
\label{eq:CConstrPrBoth}
\text{CVaR}_{1-\beta}^{s \sim \mathbb{P}}\left[\max_{\substack{t\in \mc{T},\\d\in \mc{D}}}\left(g_1(Y,s,v), g_2(Y,s,p)\right)
\right] \leq 0.
\end{align}
Note that this formulation constrains the worst case violation in terms of VCC limits and power constraints jointly and avoids introducing additional conservatism which would be the case if one added another independent CVaR constraint.

\subsection{Objective function}
The objective is a piece-wise linear function of the VCC limits and can incorporate different time-varying grid signals (carbon intensities, carbon free energy score, electricity prices, DR signals, etc.) as well as infrastructure related costs (peak power, stand-by machine cost etc.). We provide a summary of different cost components which are relevant for data center load shifting and have been used in the literature in Table~\ref{tb:CostElements}. Here, we use a similar objective function to Google's real implementation in~\cite{radovanovic2023carbon}, considering carbon footprint and daily peak power as a function of VCCs:
\begin{align} \label{eq:Cost}
f(v) = \sum_{t\in \mc{T}} \sum_{d\in \mc{D}}  \rho^{\text{carb}}_{t,d}~v_{t,d} + \sum_{d\in \mc{D}} \rho^{\text{in}}_d \, \|\text{col}(v_{t,d})_{t\in\mc{T}}\|_{\infty},
\end{align}

where $\rho^{\text{carb}}_{t,d}$ is a metric for the generated carbon footprint~\cite{electricitymaps2024carbon} and $\rho_d^{\text{in}}$ is associated with infrastructure costs driven by a cluster's peak power consumption. Peak power is used as an important signal to guide when to increase capacity by adding more machines in a cluster. The carbon impact of utilizing computing power is modeled using a linear relationship, as described in~\cite{xu2020managing, lindberg2022using}. As predictions for time-varying grid signals can be inaccurate or change rapidly, so re-optimizing throughout the day is necessary for better performance and to participate in DR schemes. 
The parameter $\rho^{\text{carb}}_{t,d}$ ideally represents how 1 MW increase in load at time $t$ and cluster $d$ affects the system's total carbon emissions. However, it is important to note that carbon intensity data is generally not available in real time. Instead, estimates must be made using the absolute values provided by~\cite{electricitymaps2024carbon}.


\begin{table}[tb]\renewcommand{\arraystretch}{1.4} 
\centering\caption{Metrics for load‑shifting decisions in data centers.\label{tb:CostElements}}
\begin{tabular}{@{}>{}l>{}c>{}c@{}} 
\toprule 
 Description  & & Publication \\ \midrule
 Generator cost && ~\cite{chen2016robust}\\
 Dispatch/power cost & & \cite{dvorkin2024agent, nadalizadeh2021greenpacker}\\
Carbon intensity (local, average, marginal) &&  \cite{breukelman2024carbon, james2019low}\\
Migration of load (delay, network charges) & &\cite{breukelman2024carbon, james2019low, narayanan2017right}\\
Fragmentation cost &&  \cite{chen2016robust}\\
Max power or peak load &  & \cite{breukelman2024carbon}\\
 & & \\
\end{tabular}
\end{table}

\subsection{Demand response}
Google is offering demand response (DR) services in DCs across the globe~\cite{mehra2024using}. During extreme weather events in Nebraska Google has already successfully supported the local utility in avoiding a blackout by reducing power consumption~\cite{mehra2024using}. Beyond monetary incentives, DC operators may need to offer flexibility to utilities in regions with strained power grids if they wish to secure building permissions and sufficient power capacity in the future. Typically, the local distribution system operator will provide a demand-response signal day-ahead, and Google would declare compute capacity in MW that can be dropped if requested. Thus, the price signal is available in advance but if and when the demand response event happens during the next day is uncertain. We model such events using scenarios $s_{k,\text{DR}}$ of a new class without spatial or temporal flexibility. Thus, any DR request will be placed immediately in the given cluster. All constraints \eqref{eq:TSconstr}-\eqref{eq:CConstrPrBoth} remain unchanged, but now apply to the extended class set $\tilde{\mc{C}}:= \{1,\dots,C,\text{DR}\}$, which includes the additional DR class. The advantage of treating DR events using a stochastic model is that the job completion requirements still hold in expectation as defined by the CVaR constraint. 
\begin{rmk}
Without available data, capacity for DR events must be reserved across all time steps, effectively leading to a uniform robustification. However, when historical data or prior assumptions about DR events is accessible, capacity can be strategically reserved only for periods with a high probability of a DR event occurring, thereby  reducing conservatism and improving cost performance.
\end{rmk}



\subsection{Stochastic scheduling problem}
Overall, the problem of co-designing the optimal scheduling strategy and VCCs is a stochastic program of the form
\begin{subequations}
\label{eq:CVARprob}
\begin{align}\label{eq:CVARprob_cost}
		\min_{y,v} & \; f(v) \\
		\textrm{s.t.} 
		&\; \text{CVaR}_{1-\beta}^{s \sim \mathbb{P}}[F(y,v,s)]  \leq  0, \label{eq:F}\\
		&  \quad~\eqref{eq:TSconstr},\ ~\eqref{eq:SpatTempConst}, \ ~\eqref{eq:VCC_constraint_with_base_load},\qquad \qquad \qquad \forall k,c,t,d, 
\end{align}
\end{subequations}
where $y = \text{col}(y_{r})_{r=0}^{KCTD}$, $y_r = Y_{k,c,t,d}$ with $r = K(c-1) + k + KC(t-1) + KCT(d-1)$ is a vectorized version of the load schedule, $Y$, which is more suitable as an optimization variable. The constraint~\eqref{eq:F} is a reformulation of~\eqref{eq:CConstrPr} in terms of $y$ with $F$ defined as
\begin{equation}
F(y,v,s):= \max_{t\in\mathcal{T},\, d\in\mathcal{D}}\ y^\top\! A_{td}\ s + v^\top b_{td} \,,
\end{equation}
where the matrix $A_{td} \in \R^{KCTD \times KC}$ has ones in its $(kctd,kc)$ entries and zeroes elsewhere, while the vector $b_{td} \in \R^{TD}$ has its $td$ entries set to $-1$ and zero elsewhere. An example of a load schedule is shown in Figure~\ref{fig:VCC}. Note that we omit adding the constraints for the power limit~\eqref{eq:ConstrMaxPow} and ramping limit~\eqref{eq:RampLim} in the scheduling problem~\eqref{eq:CVARprob} and the subsequent theoretical derivations for improved clarity. However, the derivations with constraints \eqref{eq:ConstrMaxPow}-\eqref{eq:CConstrPrBoth} included would follow the same structure.  

\begin{rmk}
\label{rem:epi}
The stochastic problem in \eqref{eq:CVARprob_cost} could be formulated only in terms of scheduling strategy $y$ and the random load profile $s$. 
 The epigraph reformulation we chose through constraint~\eqref{eq:CConstrPr} gives an explicit CVaR bound $v_{t,d}$ for each cluster $d$ and time $t$ for the random aggregate load profile $s_{k,c}$. As done by Google~\cite{radovanovic2023carbon}, the virtual capacities $v_{t,d}$ can be enforced as hard constraints on each cluster's executable load in each cluster.
{\hfill $\square$}
\end{rmk}

Unfortunately, the stochastic problem~\eqref{eq:CVARprob} cannot be solved directly, since the underlying unknown probability distribution $\bb{P}$ of the aggregate load $s$ is unknown. Instead, we only have access to historical data, provided as $24$-hour aggregate load samples for each flexibility class $s^i_{c}$. In the next section, we show how we leverage data-driven distributionally-robust optimization (DRO)~\cite{esfahani2017data} to solve~\eqref{eq:CVARprob}, essentially building a robust forecast into the optimization problem using the finite set of historical data.

\section{Distributionally-robust scheduling}\label{sec:DROAllocation}

Assuming that the per-class load profiles $s^i_c$ in the dataset are independent samples drawn from the unknown distribution $\mathbb P$, we can derive an empirical compute load distribution as
\begin{equation}
\label{eq:EmpDistribution}
	\hat{\mathbb{P}} = \frac{1}{N}\sum_{i=1}^{N} \delta_{s^i}\ ,
\end{equation}
where $\delta_{s^i}$ denotes the Dirac distribution centred at the vector $s^i\in \mathbb{R}^{KC}$, namely, the $i^{\text{th}}$ sample of per-class aggregate load obtained from the historical data. This empirical aggregate load distribution offers an approximate representation of the distribution of loads expected for the next day.

One could solve problem~\eqref{eq:CVARprob} with respect to the empirical distribution $\hat{\mathbb{P}}$ in place of the true distribution $\mathbb{P}$, obtaining the so called Sample Average Approximation (SAA)~\cite{kleywegt2002sample}.
While simple to implement, the SAA relies on the fact that the distribution $\hat{\mathbb{P}}$ built from the historical data can be used as an accurate forecast of the distribution of the future aggregate load. When this is not the case, the SAA may lead to poor out-of-sample performance and cannot provide tight probabilistic guarantees if a limited amount of data is available or if there is a mismatch between the approximate and true aggregate load distribution. 

In practice, the probability distribution of the aggregate load can exhibit significant variations across different clusters and through seasonal changes~\cite{subirats2015assessing}. Moreover, the load on specific days, such as public holidays or close to project deadlines, can deviate substantially from historical patterns, making predictions particularly challenging~\cite{subirats2015assessing}. Consequently, the empirical distribution~\eqref{eq:EmpDistribution} may not fully capture the nuances of future load distributions. This discrepancy leads to a potential shift, the infamous distribution shift, between the empirical distribution, which informs the derivation of the optimal load schedule, and the actual future load distribution encountered.

To address this uncertainty, we formulate a distributionally-robust version of~\eqref{eq:CVARprob} where the optimization is carried out against the worst-case distribution in a ``neighboorhood'' of the empirical distribution $\hat{\mathbb P}$, commonly known as \textit{ambiguity set}. Formally, we define this ambiguity set as
%
\begin{align} \label{eq:Ambiguity}
  \mc{B}^\varepsilon := \left\{ \bb{Q}\in\mc{P}_1(\mc{S})~|~ d_W\left(\hat{\bb{P}}, \bb{Q}\right) \leq \varepsilon \right\} \ ,
\end{align}
where $\mc{P}_1(\mc{S})$ is the set of Borel probability measures with finite first moment, and $d_W\big(\hat{\bb{P}}, \bb{Q}\big)$ is the so-called Wasserstein distance, given in Definition~\ref{dfn:WasserMet}, between the probability distributions $\bb{Q}$ and $\hat{\bb{P}}$~\cite{villani2009optimal}. Intuitively,~\eqref{eq:Ambiguity} describes the set of distributions that are within a radius $\varepsilon$, as measured by the Wasserstein metric $d_W$ of the empirical distribution $\hat{\mathbb{P}}$. Ambiguity sets based on the Wasserstein distance are expressive and particularly well-suited for modeling and robustifying against so-called black-swan events, namely, rare and unpredictable outlier events with extreme impact\footnote{For instance, an extremely large compute load that, under classical placement schemes, would lead to execution delays for time-sensitive jobs.}~\cite{taleb2010black}. 

The radius $\varepsilon$ of the ambiguity set in~\eqref{eq:CVARprob_robust_C} shall be chosen such that the ambiguity set is large enough to contain the true (unknown) future load distribution. It  can be regarded as a tuning knob that allows to trade-off between performance (cost reduction) and probabilistic constraint satisfaction. Namely, for $\varepsilon=0$ we recover the SAA formulation that is not robust against distribution shifts whereas for large $\varepsilon$ the solution to~\eqref{eq:CVARprob_robust} will be robust but possibly conservative. In practice, a suitable radius $\varepsilon$ can be obtained by analyzing the available data, e.g., by cross-validation.

With these definitions in place, we can formulate the distributionally-robust version of~\eqref{eq:CVARprob} as
\begin{subequations}%
\label{eq:CVARprob_robust}
\begin{align}
		\min_{y,v} \quad & f(v) \\
		\textrm{s.t.} \quad
		& \sup_{\mathbb{Q} \in \mathcal{B}^\varepsilon} \text{CVaR}_{1-\beta}^{s \sim {\mathbb{Q}}}[F(y,v,s)] \leq  0, \label{eq:CVARprob_robust_C}\\
            &  \quad~\eqref{eq:TSconstr},\ ~\eqref{eq:SpatTempConst}, \ ~\eqref{eq:VCC_constraint_with_base_load},\qquad \qquad \qquad \quad \forall t,d,c,k.
\end{align}
\end{subequations}
This approach enables us to achieve robust (probabilistic) constraint satisfaction while providing a probabilistic guarantee on the realized cost. 
Furthermore, the coupled optimization of VCCs and scheduling $Y$, and the use of an ambiguity set centered on the empirical distribution of the available data distribution, allow us to exploit the temporal and spatial correlations that exist across job classes.

Remarkably, the worst-case constraint over the set of distribution in the ambiguity set that appears in problem~\eqref{eq:CVARprob_robust} admits a tractable reformulation as a linear program (LP) that depends on the observed samples $s^i \in\mathcal{S}$, with $\mathcal{S}$ being an a priori known support set for $\mathbb{P}$ defined here as
\begin{align}
\mc{S}:=\{s\in\R^{KC}~|~Gs\leq h\}.
\end{align}
The matrix $G\in\mathbb{R}^{g\times KC}$ and vector $h\in\mathbb{R}^{g}$ encode prior information on the random aggregate load $s$, for example, the fact that job volumes can only be positive and that there exists a finite upper bound on the total aggregate load. A solution to the distributionally robust problem~\eqref{eq:CVARprob_robust}, i.e., the optimal schedule $y^*$ and VCC $v^*$, can thus be obtained by solving an LP as shown in the following proposition.

\begin{figure*}[ht]
\centering 
\includegraphics[width=2\columnwidth]{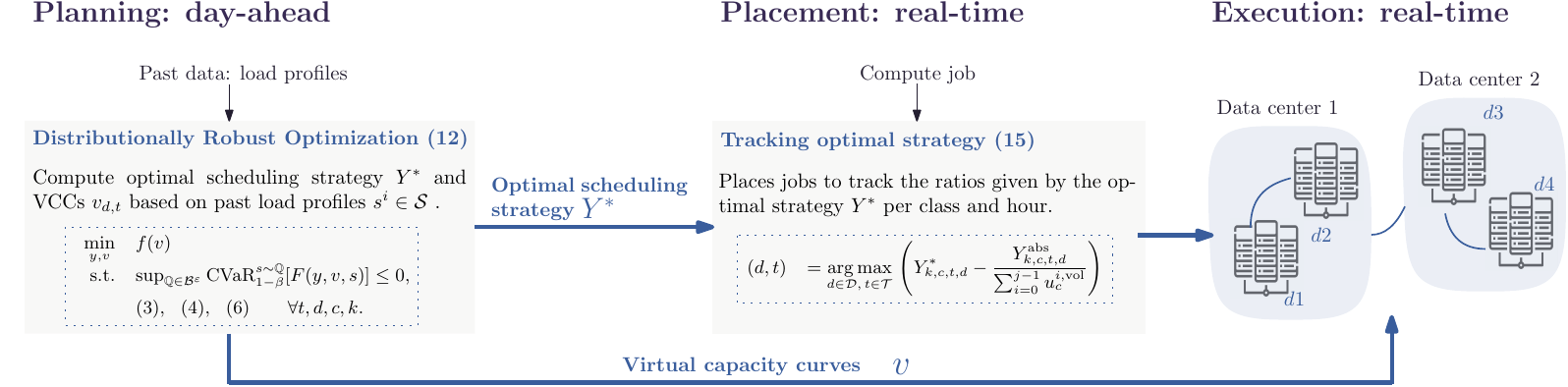}
\vspace{5pt}
\caption{Schematic of the two layered control approach separated into day-ahead planning and real-time execution.} \label{fig:PipelineSchematic}
\end{figure*}

\begin{prp}\label{prp:DRO}
    Assume that $s\in\mc{S}$ then, the DRO problem~\eqref{eq:CVARprob_robust} can be reformulated as an LP\footnote{The infinity norm term in the cost function~\eqref{eq:Cost} can be reformulated in LP form, but we omit it here to focus on the constraint reformulation.}
    \begin{equation}\label{eq:DRO_reformulation}
        \scalemath{0.94}{\begin{aligned}
            \min_{\substack{y,v,q\\p^i,\lambda, \eta_{itd}}}& f(v) \\
    	\textrm{s.t.} \
            & \, \lambda\varepsilon + \frac{1}{N}\sum_{i=1}^{N} p^{i} \leq q \beta \\
    	& \!\left[v^{\top}\!b_{td} + q + 
            \left(y^{\top} \!A_{td} - \eta_{itd}^\top G \right) s^i + \eta_{itd}^\top h \right]_+\!\leq p^i \\
    	& \!\left\| y^\top\!\! A_{td} - \eta_{itd}^\top  G\right\|_\infty \leq \lambda\\
            & q \in \R,\  \eta_{itd} \geq 0, \ \lambda\geq 0\\
            &~\eqref{eq:TSconstr},\ ~\eqref{eq:SpatTempConst}, \ ~\eqref{eq:VCC_constraint_with_base_load}, \\
            & \forall\  t\in\mathcal{T},\ d\in\mathcal{D},\ c\in\mathcal{C},\ k\in\mathcal{H},\ i\in\{1,\dots,N\}
        \end{aligned}}
    \end{equation}
where $q\in\mathbb{R}\,$, $\lambda\in\mathbb{R}\,$, $p^i\in\mathbb{R}\,$, and $\eta_{idt}\in\mathbb{R}^{g}$ are auxiliary variables.
\end{prp}

\begin{proof}
    The proof follows directly from Proposition V.1. in~\cite{hota2019data} and is omitted here due to space limitations.
\end{proof}
The number of decision variables scales quadratically with $T$, linearly with $C$ and $D$. Further, the formulation naturally lends itself to decomposition techniques based on spatial constraints of jobs.

Next, we show that any solution to the DRO reformulation is a feasible point of problem~\eqref{eq:CVARprob}.
\begin{prp}\label{prp:}
Assume that $\varepsilon$ is such that $\bb{P}\in \mathcal{B}^\varepsilon$, and that $(y^*, v^*)$ are component of a solution $(y^*,v^*,q^*,p^{i*},\lambda^*,\nu^*)$ to the LP~\eqref{eq:DRO_reformulation} in Proposition~\ref{prp:DRO}. Then, $y^*$ and $v^*$ are a feasible solution to the original stochastic optimization problem~\eqref{eq:CVARprob}.
\end{prp}%
\begin{proof}
The proof follows directly from the definition of the worst-case distribution through the ambiguity set in~\eqref{eq:Ambiguity}. If~\eqref{eq:DRO_reformulation} is feasible for all distributions in the set $\mc{B}^\varepsilon$, then it is also feasible for true unknown distribution~$\mathbb{P}$ as the ambiguity set contains $\mathbb{P}$ by assumption.
\end{proof}

This statement shows that if there exists a solution to the LP in Proposition~\ref{prp:DRO}, then the constraint satisfaction is guaranteed for all distributions inside the ambiguity set $\mathcal{B}^\varepsilon$.  In other words, for any feasible scheduling strategy (thus, also for the optimal one), given a new aggregate load profile $s^{\text{new}}$ drawn from a distribution inside the ambiguity set, the following holds:
\begin{enumerate}
\item[(i)] With high probability\footnote{As specified by the CVaR in \eqref{eq:CVARprob_robust_C}, tuned by adjusting parameter $\beta$.}, the total compute load deployed on cluster $d$ at time $t$ will not exceed the planned operational capacity $v^*_{t,d}$;
\item[(ii)] 
With high probability, the carbon footprint and peak load cost of the data center network over the $24$-hour horizon is upper-bounded by $f(v^*)$.
\end{enumerate}

The proposed method differs from the approach in~\cite{radovanovic2023carbon} in the formulation of the constraints and in how robustness is introduced in the optimization problem. In \cite{radovanovic2023carbon}, robustness is obtained using the $97^{th}$ percentile bound of the forecasted compute load. In contrast, we achieve robustness by formulating the DRO~\eqref{eq:CVARprob}. The level of robustness is controlled by the Wasserstein ambiguity set radius $\varepsilon$, while the risk level can be adjusted by varying the $\beta$ parameter of the CVaR-based constraint. Compared to the approach in~\cite{radovanovic2023carbon}, our proposed co-design of VCCs and scheduling enables directly exploiting the correlation of load profiles between classes which emerges from the data, as captured by the empirical compute load distribution $\hat{\mathbb{P}}$.


\section{Real-time job placement}\label{sec:RTPlacement}

In the real application of job placement, the aggregate load of each class $c$ consists of discrete compute jobs submitted continuously over the planning horizon. Thus, we can not perfectly implement the scheduling strategy $Y^*$ computed in~\eqref{eq:DRO_reformulation}. Instead, in this section we present a real-time placement scheme which uses $Y^*$ as a reference signal and tries to track it with every incoming job. 

In fact, in real-time operation, the scheduler needs to make hundreds of thousands of placement decisions per second~\cite{tirmazi2020borg} making it imperative that the placement algorithm is computationally inexpensive, allowing jobs to flow immediately to the available compute resources. To achieve this, we designed our scheme to operate on two time-scales:
\begin{enumerate}
\item Day-ahead planning: DRO problem takes into account past aggregate load data, future load predictions, as well as the constraints of jobs and data centers. The output of the DRO problem is the optimal scheduling strategy $Y^*$ which acts as a reference signal for job placement and the VCC curve for every cluster $v_{t,d}, \; \forall t \in \mc{T}, \forall d \in \mc{D}$.

\item Real-time job placement: Places every incoming job with the goal of tracking the optimal scheduling strategy $Y^*$.

\end{enumerate}
The approach is illustrated in Figure~\ref{fig:PipelineSchematic}. The time-scale separation of the overall control architecture allows to have a complex planning problem with rigorous theoretical guarantees while having a simple real-time placement which can handle a continuous flow of incoming compute jobs. 


We point out that in the DRO problem used for day-ahead planning it is assumed that the continuous aggregate load profile can be split into arbitrary small fractions given by the scheduling strategy $Y_{k,c,t,d}$. However, in reality the aggregate load $s_{k,c}$ of class $c$ at time $k$ is the sum of compute volumes of discrete jobs $s_{k,c}= \sum_{j} u_c^{j, \text{vol}}$. Thus, in practice, the extent to which the optimal scheduling strategy $Y^*$ can be approximated with discrete jobs depends on their size, number and runtime. In case of unexpected real-time changes, such as machine failure, the VCC limits will be recomputed the next day reducing the affected cluster’s capacity and redistributing load. If one would like the approach to handle cluster or machine failures in real-time, then a protection mechanism could be implemented in the job placement algorithm.

More details on application specific considerations and extensions are given in Subsection~\ref{sbs:ApplCons}.


In real-time operation, a job $u$ of class $c$ arrives, and we want to place it in cluster $d$ and time slot $t$ which is furthest away from fulfilling the optimal load fraction $Y^*$. Thus, for the j$^{\text{th}}$ job of volume $u_c^{j,\text{vol}}$ arriving in hour $k$, we choose the placement tuple $(d,t)$ for which the difference between $Y^*$ and the current load ratios $\frac{Y^{\text{abs}}_{k,c,t,d}}{\sum_{i=0}^{j-1} u_c^{i,\text{vol}}}$ is maximal, expressed in the following:
\begin{align}
\label{eq:alloc_rule}
(d,t) &=\displaystyle \argmax_{d\in \mc{D},\,t\in \mc{T}} \,\bigg(\underbrace{Y^*_{k,c,t,d}}_{\substack{\text{Optimal}\\\text{fraction}}}- \underbrace{\frac{Y^{\text{abs}}_{k,c,t,d}}{\sum_{i=0}^{j-1} u_c^{i,\text{vol}}}}_{\substack{\text{Current}\\\text{fraction}}}\bigg)
\end{align}
where $Y_{k,c,t,d}^*\in \R$ is the optimal scheduling strategy for jobs of class $c$ submitted at time $k$ and $u_c^{i, \text{vol}}$ is the compute volume of job $i$. The tensor $Y^{\text{abs}}_{k,c,t,d}\in \R$ is the absolute compute volume already placed for execution in cluster $d$ at time $t$. It is initialized at the beginning of every hour $k$ before the first job arrives, i.e., when $j = 0 $, $Y^{\text{abs}}_{k,c,t,d} = 0$. In practice, a job placed at a cluster $d$ for execution at time $t$ is stored in a cloud until the optimal time of execution, i.e, until $k=t$. A cluster-level operating system (known as Borg at Google \cite{verma2015large, tirmazi2020borg}) handles placements to available virtual machines. 

The computational complexity of our real-time job placement scheme~\eqref{eq:alloc_rule} is determined by finding the maximum value in a vector/tensor. This operation has a complexity of $O(D×C)$, where $D$ is the number of clusters, and $C$ is the number of job classes.

%
%


\section{Implementation and Numerics}\label{sec:Numerics}

In this section, we discuss relevant extensions to our model tailored to real-world application and present illustrative simulation studies to showcase the main features and principles of our DRO scheduling strategy. To keep the focus clear and illustrative, we consider sinusoidal cost function parameters as a representative example of time-varying grid signals. Further, we point out that the simulations are presented for 24hrs of execution time instead of $T = K + \max h_c = 34$hrs as new VCC limits get pushed every day. We assume that remaining jobs will fit within the VCC limits of the next day which is consistent with the approach in~\cite{radovanovic2023carbon}. 

\subsection{Application specific considerations and extensions}\label{sbs:ApplCons}
\begin{enumerate}
    \item \textbf{Job runtimes:} We currently assume homogeneous job runtimes of one time step. This is consistent with the industry practice of splitting large compute jobs into small parallelizable tasks. To handle longer runtimes, we can introduce additional job classes and constraints to ensure time continuity of the scheduling strategy.
    \item \textbf{Cross-resource requirements:} Compute jobs require a variety of resources on machines (CPU, memory, disk, etc.). The DRO problem~\eqref{eq:CVARprob_robust} can be formulated taking into account job's usage across all resource dimensions by considering vectorized VCC limits per cluster.
    \item \textbf{Capacity constraints:} The VCCs $v_{t,d}$ sent to clusters can be used in two ways: (i) As information for the cluster-level operating system (known as Borg at Google \cite{verma2015large, tirmazi2020borg}) on the CVaR bound of the daily load profile; (ii) As a hard capacity constraint for the placement algorithm. The implementation by Google in~\cite{radovanovic2023carbon} incorporates both (i) and (ii). We choose approach (i) In Section~\ref{sbs:SimSetup} through Section~\ref{subsec:CVArParamSim} as it aligns with the theoretical derivations in Section~\ref{sec:ProblemSetting} and~\ref{sec:DROAllocation}. For Section~\ref{sec:PracticalImp} we use approach (ii) which is closer to the practical implementation. 
    \item \textbf{Data handling:} 
    Samples $s^i$ used to define~\eqref{eq:EmpDistribution} and to solve~\eqref{eq:DRO_reformulation} can be (i) past per-class aggregate load profiles, which is simple but ignores correlations like day-of-week effects, or (ii) samples can be obtained from a calibrated stochastic predictor using large datasets, capturing side information and correlations.
    
   
    \item \textbf{Computational complexity:} The resulting DRO problem~\eqref{eq:DRO_reformulation} is an LP, thus large amounts of data and different classes can be included without jeopardizing solvability within one hour. Nevertheless, a data pre-processing pipeline would be essential to handle the massive amounts of data, their anomaly detection and missing data.
    \item \textbf{Receding-horizon implementation and feedback:} As load predictions and grid signals may update throughout the day, and the state of the network may change, recomputing the DRO problem~\eqref{eq:CVARprob_robust} every hour, including the system state, and applying it in a receding-horizon manner is expected to yield performance gains in practice.
\end{enumerate}


\subsection{Simulation set-up and processing of data}\label{sbs:SimSetup}

We consider a network of computing clusters $\mc{D} = \bb{Z}_{4}$ consisting of 2 data centers with two clusters each, as illustrated in the right-hand side of Figure~\ref{fig:PipelineSchematic}. For exposition purposes, we consider a simplified setup in which the maximum delay time is $h_c = 10$ hrs for all compute jobs. However, we stress that our approach generally accommodates any number of time flexibility classes. This setup results in a total of seven job classes $\mc{C} = \bb{Z}_{7}$: four cluster-bound classes $\mc{D}_{c}=d ,\forall c=d\in\{1,2,3,4\}$, i.e., class one is bound to cluster one, two data center-bound classes $\mc{D}_{5} = \{1,2\}$, $\mc{D}_{6} = \{3,4\}$, and one globally flexible class, $\mc{D}_{7} =\mc{D}$.

We use 60 normalized daily compute usage shapes from randomly selected clusters in the Google fleet.  
The LP in~\eqref{eq:DRO_reformulation} is solved in Python using Gurobi~\cite{gurobi} with 60 training load samples (80\% of the total dataset). We use the compute load shapes from the training data set directly as individual samples $s^i$ in the LP given in~\eqref{eq:DRO_reformulation}. The numerical case studies are conducted with 15 previously unobserved validation samples (20\% of the dataset). The aggregate compute load curves after allocation are  generated by multiplying the optimal scheduling strategy $Y^*$ with the unobserved training samples. The cost function parameters in~\eqref{eq:Cost} are generated using sinusoids with a phase shift for different clusters to emulate variations in time-varying grid signals by location. All simulations are solved for CVaR level $\beta = 0.2$ and ambiguity set radius $\varepsilon = 8 \cdot 10^{-3}$ if not indicated otherwise.



\subsection{Day-ahead: Training and validation scenarios}\label{sbs:DayAhead}

In Figure~\ref{fig:TrainValComparison} we plot the output of the DRO day-ahead planning problem~\eqref{eq:CVARprob_robust}. Specifically, we show how the optimal load strategy $Y_{k,c,t,d}^*$ would distribute the aggregate load of the training and validation scenarios $s^i$ across time $t\in \mc{T}$ and clusters $d\in \mc{D}$. We observe that even for unobserved aggregate load profiles in the validation set, load is shifted successfully and stays almost always within VCC bounds as given by the probabilistic constraints. We make the following observations:
\begin{figure}
\centering 
\includegraphics[width=\columnwidth]{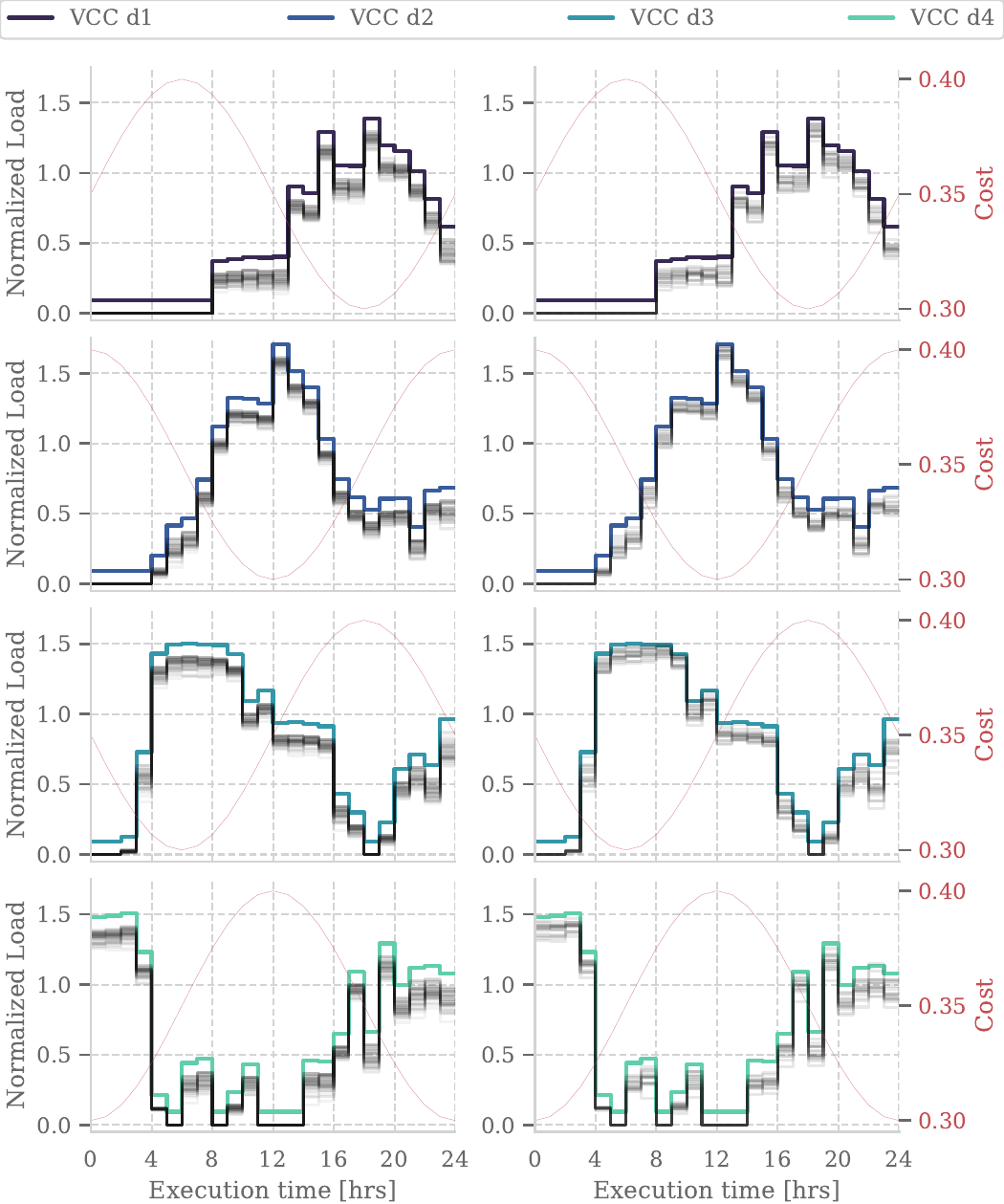}
\caption{Comparing load profiles under the optimal schedule $Y_{k,c,t,d}^*\cdot  s_{k,c}^i, \forall t\in \mc{T}, d\in \mc{D}$ for 60 training scenarios $s^i_{\text{train}}, i\in \bb{Z}_{60}$ (left) and 15 validation scenarios $s^i_{\text{val}}, i\in \bb{Z}_{15}$  (right).} \label{fig:TrainValComparison}
\end{figure}

\begin{itemize}
    \item The robust scheduling strategy, $Y^*_{k,c,t,d}$, leverages the spatial range and temporal flexibility of loads to minimize the overall cost. For example, minimal load is executed during the first 8 hours of the day at location $d=1$, where costs are considerably higher.
    \item For all the training scenarios, the optimal robust schedule results in loads slightly below the VCC capacity, $v_{t,d}$, at all times and across all clusters.
    \item For some of the validation scenarios, the optimal robust strategy results in load profiles that slightly exceed the VCC limits, $v_{t,d}$, for example for hours 8 to 10 at cluster 3. This is expected due to the probabilistic nature of the CVaR constraint in~\eqref{eq:CConstrPr}.
\end{itemize}

\subsection{Greedy vs planning }
\begin{figure}
\centering 
\includegraphics[width=\columnwidth]{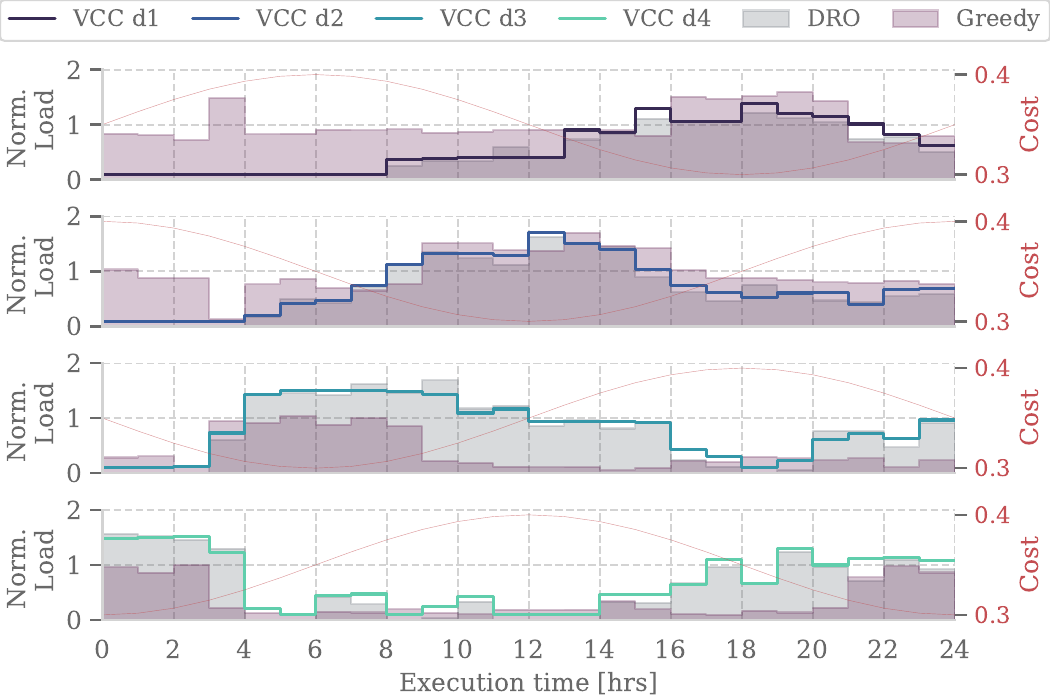}
\caption{Comparing the running load of the DRO and greedy policy over one day.\label{fig:LoadComparison}} 
\end{figure}

In this study, we compare the optimal scheduling strategy resulting from our proposed DRO approach with a basic greedy approach that places every incoming job at hour $k$ to the ``cheapest" cluster within the job classes' spatial range $d\in \mc{D}_{c}$ and with available (cluster) machine capacity. The load profiles over one day for both policies are shown in Figure~\ref{fig:LoadComparison}. Table~\ref{tb:Relative Cost} shows the cost across all 15 validation scenarios compared to an ideal placement that has a perfect forecast of the next day's aggregate load $s_{k,c}$. 
\begin{table}
\centering\caption{Comparing the cost of the DRO and greedy policy to the optimal cost with perfect forecast.}\label{tb:Relative Cost}
\begin{tabular}{@{}p{2.8cm}lll@{}}
\toprule
 Policy  & DRO & Greedy \\ \midrule
 Total cost increase:
 Mean [\%] (STD [\%] )&  $\varepsilon = 8 \cdot 10^{-3}$: $\mathbf{2.57}$ (1.05) & \multirow{2}{*}{14.05 (0.63)} \\
 & $\varepsilon = 5 \cdot 10^{-2}$:  $8.86$ (1.00)& \\
\end{tabular}
\end{table}
Our analysis reveals several key findings:
\begin{itemize}
    \item The DRO scheduling strategy shifts job execution across clusters and time to exploit low cost hours. For example, in the first 10 hours the compute load executed in $d\in \{1,2\}$ is nearly zero as load is shifted to later hours.
    \item The costs of the DRO strategy are just 2.57 \% higher than those of the policy with perfect forecast for $\varepsilon = 8 \cdot 10^{-3}$ and 8.86 \% higher for $\varepsilon = 5 \cdot 10^{-2}$. For both robustness levels, it significantly reduces cost compared to the greedy policy, as shown in Table~\ref{tb:Relative Cost}.
\end{itemize}

This study shows that the DRO approach significantly reduces operational costs compared to a simplistic greedy policy while also giving guarantees on constraint satisfaction.
%

\subsection{Influence of CVaR constraint parameters \texorpdfstring{$\beta$}{beta} and \texorpdfstring{$\varepsilon$}{epsilon}} \label{subsec:CVArParamSim}

\begin{figure}
\captionsetup{width=0.97\linewidth}
  \centering
  \begin{subfigure}{0.97\linewidth}\includegraphics[width=\linewidth]{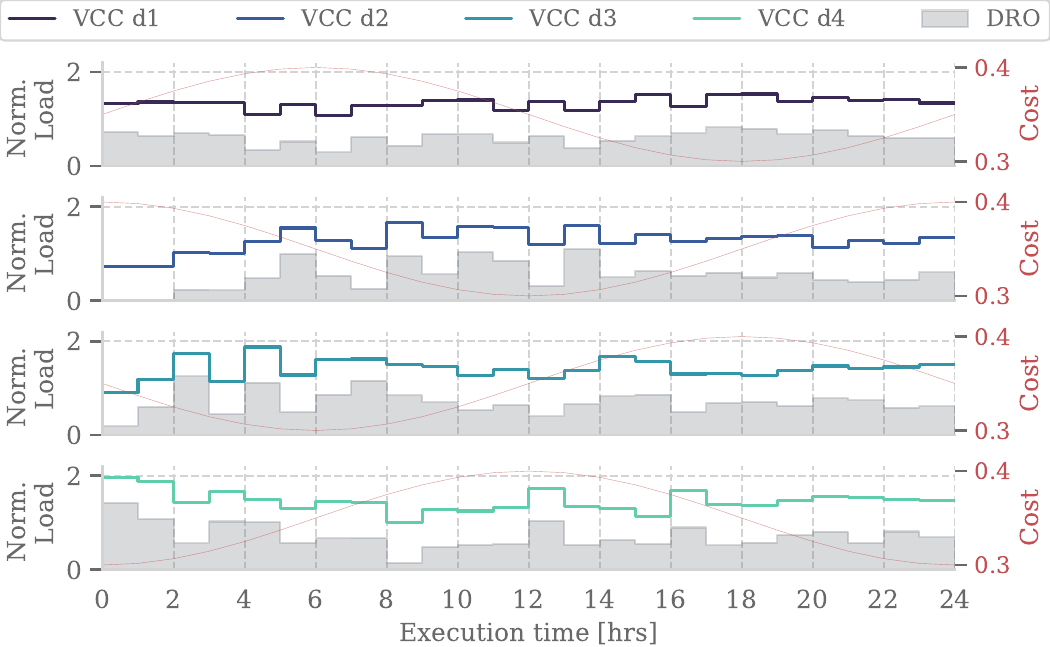}
  \caption{With $\beta = 0.02$ and \textbf{0} VCC violations. }
  \end{subfigure}\vspace{0.5em}
  \begin{subfigure}{0.97\linewidth}\includegraphics[width=\linewidth]{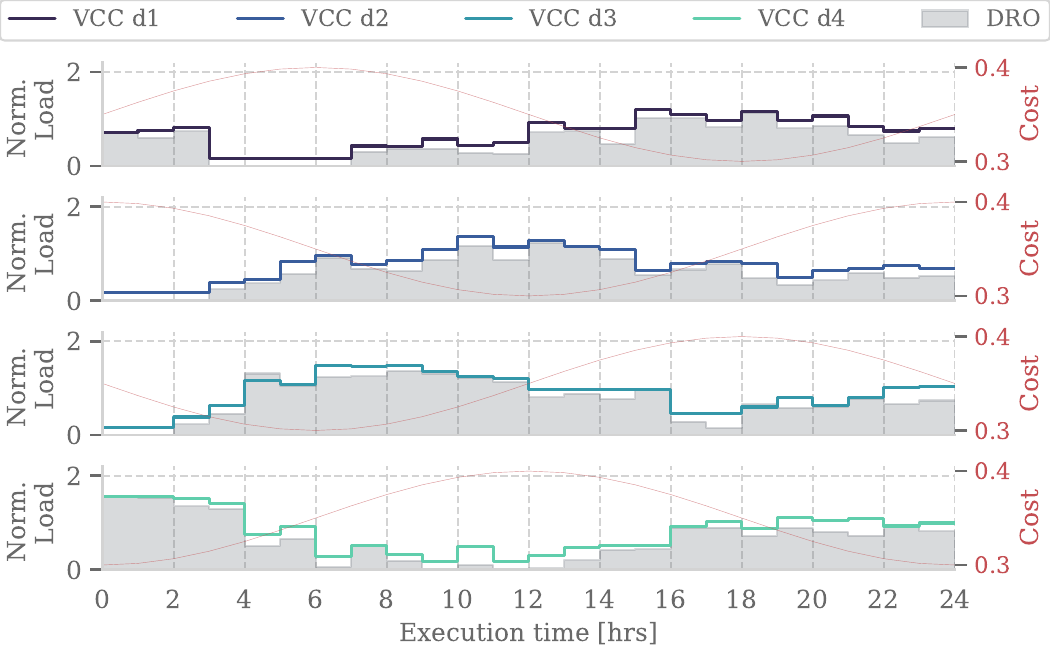}\caption{With $\beta = 0.1$ and \textbf{4} VCC violations.}
  \end{subfigure}\vspace{0.5em}
    \begin{subfigure}{0.97\linewidth}\includegraphics[width=\linewidth]{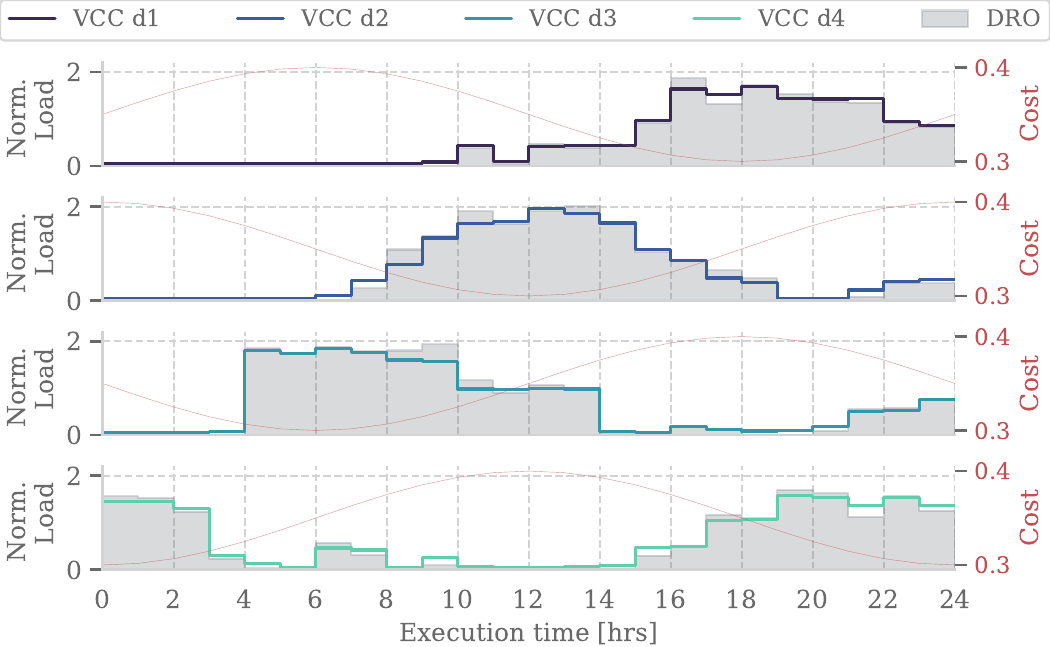}\caption{With $\beta = 0.5$ and \textbf{33} VCC violations.}
  \end{subfigure}
  \caption{Comparison of VCCs and the realised load distribution for discrete jobs submitted over the day for three different $\beta$ values.}
  \label{fig:Beta}
\end{figure}

In this section, we highlight the role of the CVaR level $\beta$ and DRO radius $\varepsilon$ influencing the CVaR constraint in~\eqref{eq:CConstrPr}. 
The parameter $\beta$ is a design parameter chosen to control the risk of constraint violations. By adapting the value of $\beta \in (0,1]$, we can model a continuous spectrum of formulations. Choosing $\beta=1$ results in a problem that requires the constraint to hold in expectation (with respect to the worst-case distribution in the ambiguity set), while choosing $\beta$ near zero results in a constraint that approaches a worst-case robust solution (with respect to the worst-case distribution in the ambiguity set). This parameter should be chosen according to the risk sensitivity of the specific application.
Figure~\ref{fig:Beta} shows the load profiles of discrete jobs over one day for increasing values of the CVaR level $\beta$. We observe that VCCs $v_{d,t}$ tighten with increasing $\beta$, corresponding to a relaxation of the probabilistic constraint against violating the VCC limits. While a value of $\beta = 0.5$ allows for significant shifting of compute load in time, thereby reducing operational costs, it may result in (cluster) level machine capacity being violated and some jobs not being executed due to insufficient capacity. However, note that in practice, the DRO would be implemented in a receding-horizon fashion, in which case it is not necessary for all jobs to be executed by the end of the finite horizon.

\begin{figure}
\centering 
\includegraphics[width=\columnwidth]{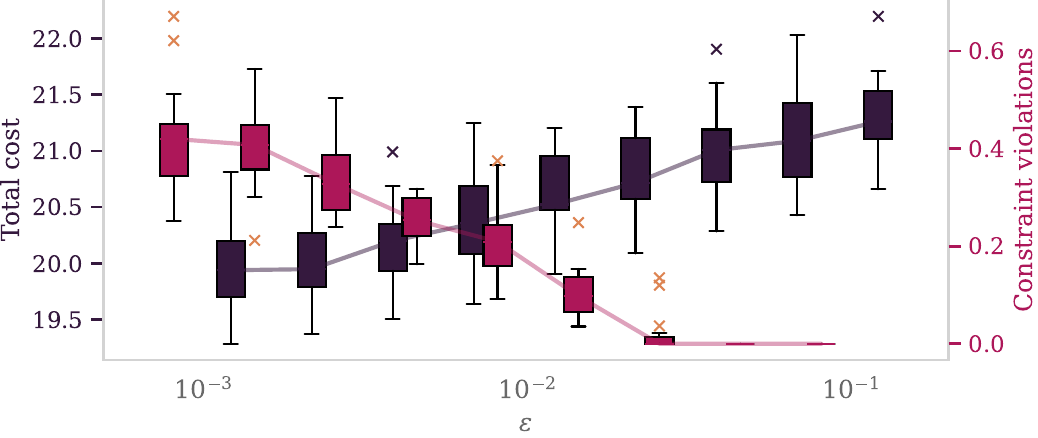}
\caption{A boxplot showing the trade-off between cost and max VCC constraint violations \eqref{eq:CConstrPr} for increasing $\varepsilon$.} \label{fig:cost_constraints_boxplot}
\end{figure}

Moreover, it is crucial to select an appropriate radius $\varepsilon$ to guarantee that the true (unknown) distribution $\mathbb{P}$ belongs to the ambiguity set $\mc{B}^\varepsilon$ in~\eqref{eq:Ambiguity}. When samples used to construct the ambiguity set center $\hat{\mathbb{P}}$ are drawn from some stationary $\mathbb{P}$, concentration results can establish confidence bounds on the Wasserstein distance between $\hat{\mathbb{P}}$ and $\mathbb{P}$ \cite{kuhn2019wasserstein}, and thus provide guidance on how to choose the radius $\varepsilon$.  However, in practical applications, these theoretical bounds often prove overly conservative. Moreover, the assumption of a stationary load-generating distribution may not hold in many real-world scenarios. The parameter $\varepsilon$ is, in practice, treated as an empirical tuning knob, at the expense of formal theoretical guarantees. A practical way of picking a value of the radius that is not too conservative is to perform a leave-one-out cross validation, using $N-1$ scenarios to build the empirical distribution $\hat{\mathbb{P}}$ and test against the left-out sample for different values of $\varepsilon$. The choice of radius can be guided by selecting a value of $\varepsilon$ that balances cost and constraint violations.
Figure~\ref{fig:cost_constraints_boxplot} shows how different values of $\varepsilon$, provide different trade-offs between performance and robustness against previously unseen future realizations of the compute loads. The problem reduces to SAA with $\varepsilon = 0$ and a robust optimization problem for $\varepsilon$ large. A choice of $\varepsilon= 10^{-2}$ achieves a good balance between performance and VCC constraint violations across all 15 validation scenarios.

\subsection{Exploiting spatial range and temporal flexibility}
\begin{figure}
\centering 
\includegraphics[width=\columnwidth]{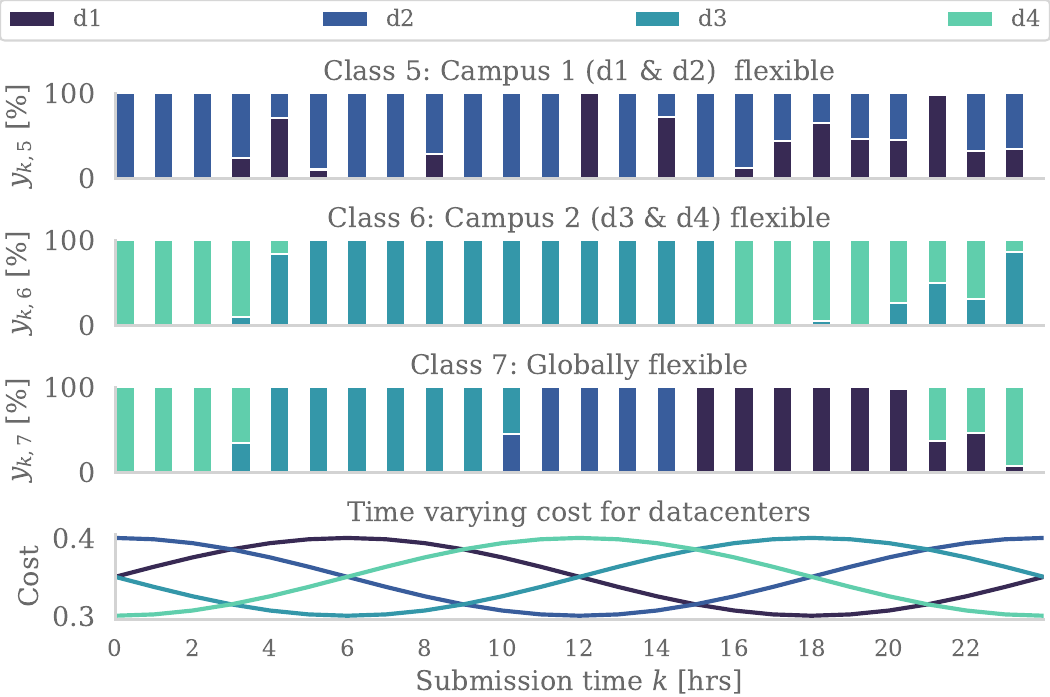}
\caption{Plotting the optimal schedule $Y^*$ for spatially flexible classes $c \in \bb{Z}_{[5:7]}$, showing the percentage of aggregate load submitted at hour $k$ that should be sent to clusters $d \in \bb{Z}_{4}$. } \label{fig:SpatialFlexibility}
\end{figure}

In this study, we aim to demonstrate that the DRO scheduling strategy utilizes the spatial range and temporal flexibility of jobs to shift them to less expensive hours and clusters. In Figure~\ref{fig:SpatialFlexibility}, we present the optimal load fractions for all spatially flexible classes: DC-flexible classes $c=4$ and $c=5$, and the globally flexible class $c=6$. The load distribution for these classes aligns with cluster-level price curves to avoid high-cost periods. For example, in the globally flexible class $c=7$, all load shifts to $d=4$ in the early hours to minimize costs.

\begin{figure}
\centering 
\includegraphics[width=\columnwidth]{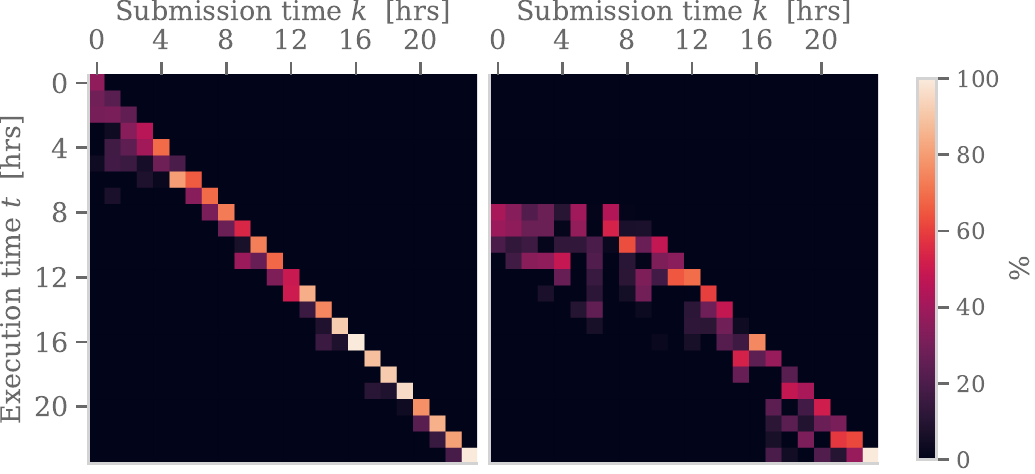}
\caption{Two heatmaps of $Y^*_{k,c,t,d=1}$ show the aggregate load percentages submitted at $k$ to be executed at $t$, with $\beta = 0.02$ (left) and $\beta = 0.2$ (right). With increased flexibility ($\beta = 0.2$) hours with high cost at $d=1$ are avoided, e.g., hours 0 to 8.} \label{fig:TemporalFlexibility}
\end{figure}

In Figure~\ref{fig:TemporalFlexibility}, we illustrate how the DRO scheme exploits temporal flexibility with the two time axes, submission and execution time. Without the second time axis (i.e., if \(t=k\)), the heatmap would show only diagonal entries, indicating 100\% execution at submission time, thereby eliminating any temporal flexibility and enforcing that all loads are processed immediately. Figure~\ref{fig:TemporalFlexibility} also highlights the effect of the tuning parameter $\beta$. For \(\beta = 0.02\), much of the load is executed immediately upon submission to meet the delay constraint of \(h_c=10\) hours. In contrast, for $\beta = 0.2$, the probabilistic constraint in~\eqref{eq:CConstrPr} is relaxed, allowing compute loads to be pushed to later hours to minimize costs.

\section{Realistic implementation with carbon cost and power constraints}\label{sec:PracticalImp}

In the real application presented by Google in~\cite{radovanovic2023carbon}, VCC curves serve as a hard constraint for resource availability and the cluster-level operating system. When the load pushed to a specific cluster exceeds its planned capacity, jobs are placed in a cluster-level queue and executed as resources become available. Note that our approach is fully agnostic to the low level control of the cluster, in the real system placement to machines in a cluster is handled by a system called Borg~\cite{tirmazi2020borg}. Here, we choose to implement a simple ``First-In-First-Out" queuing system. In this numerical study, we tested how our proposed real-time job placement scheme in~\eqref{eq:alloc_rule} handles discrete jobs. Jobs have a one-hour runtime and volume size drawn from a per-class normal distribution $\mc{N}_c(u_c^{\text{vol,mean}}, 0.1)$, with $u_c^{\text{vol,mean}}$ being the per-class compute load divided by job count.

\subsection{Queue length increases with tighter VCCs}

\begin{figure}
\centering 
\includegraphics[width=\columnwidth]{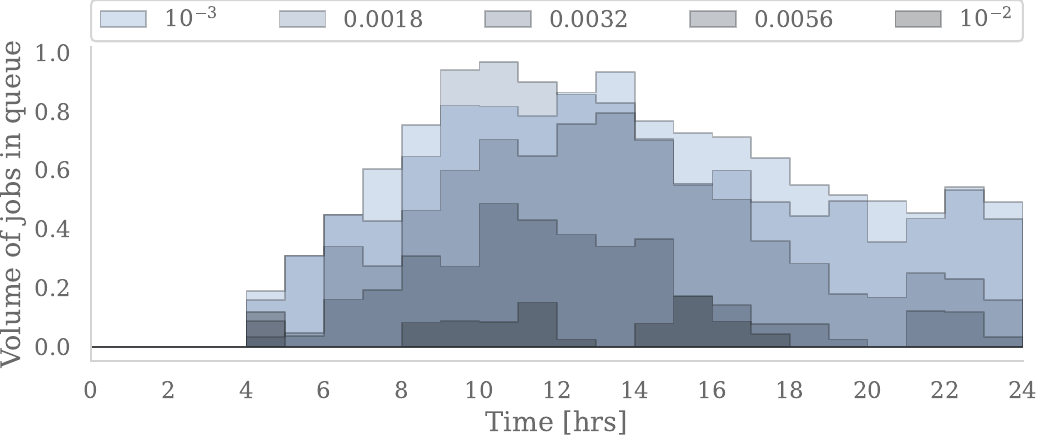}
\caption{The queue length at cluster $d=3$ for increasing $\varepsilon$.} \label{fig:queue_epsilon}
\end{figure}

In Figure~\ref{fig:queue_epsilon} we compare the resulting queue length for different radii of the ambiguity set $\varepsilon$. We make three key observations: (i) The queue length is shortest during hours 0–10, as jobs are executed immediately during low-cost periods; (ii) Queue lengths are considerably longer for lower $\varepsilon$ values, as the VCC limits become tighter and it is more likely that the scenario realization is outside the ambiguity set defined by the past data; (iii) A receding-horizon implementation that adjusts the planning policy by integrating the feedback from the queue lengths would be necessary in a real application.

\subsection{Ramping limits affect VVCs and cost}
\begin{figure}
\centering 
\includegraphics[width=\columnwidth]{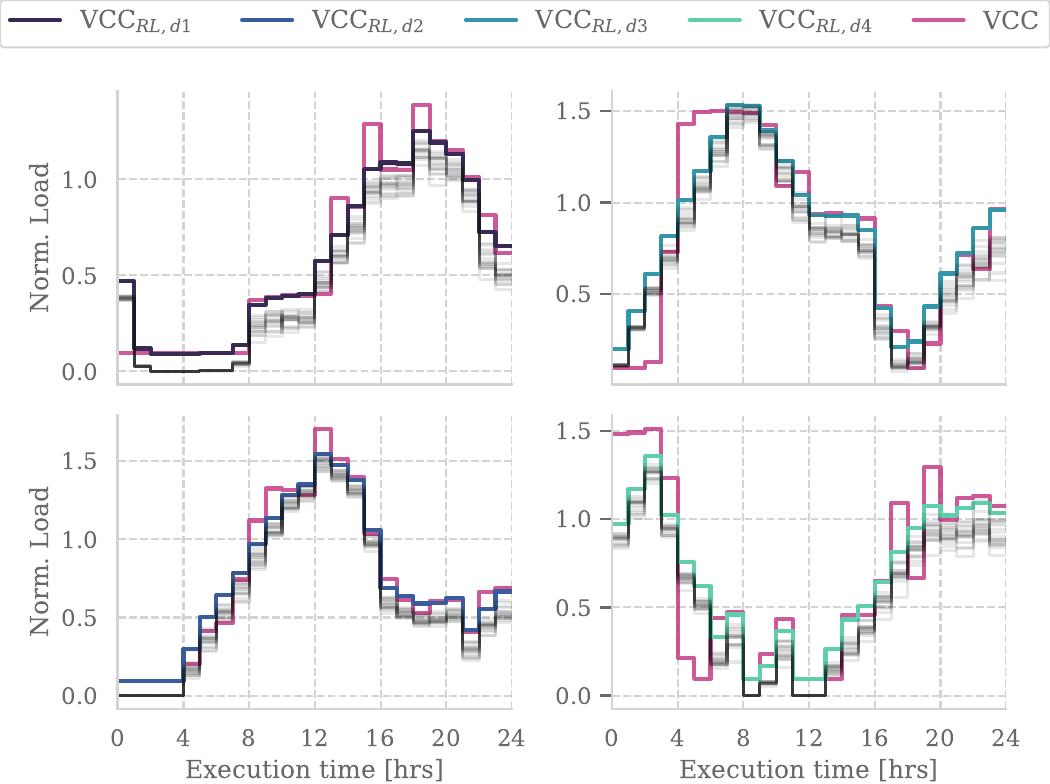}
\caption{Showing load profiles (black) for 15 validation scenarios $s^i_{\text{val}}, i\in \bb{Z}_{15}$ and capacity curves VCC$_{RL,d}$ (blue tones) under ramp constraints as defined in \eqref{eq:RampLim}, and comparing them to VCC limits without ramp constraints (magenta).} \label{fig:RampLim}
\end{figure}

We implement ramping limits, as defined in \eqref{eq:RampLim}, in the DRO problem and compare the resulting VCC limits, VCC$_{\text{RL},d}$, to the those obtained without imposing ramping limits. We present the results in Figure~\ref{fig:RampLim}. One can clearly see that the constraints induce a smoothing of the VCC limits, especially in d=3 and d=4, the major increase and drops happening between hours 3 and 6 are avoided. Here we would like to point out that it is important to enforce the constraint for the probabilistic load distribution, as defined through constraints \eqref{eq:RampLim} and \eqref{eq:g2}. If we only enforced them in terms of the deterministic VCC limit, then it might happen that the true realized load does not fulfill the ramping limits. As for instance in certain hours the load may be much lower than the VCC constraint and then jump up suddenly up to the constraint boundary, violating the ramping limit albeit the VCC limits satisfying it. This highlights once again the advantage of using a probabilistic data-driven modeling approach, as the one we present here.

\subsection{Cost function elements and VCC effects}

\begin{figure}
\centering 
\includegraphics[width=\columnwidth]{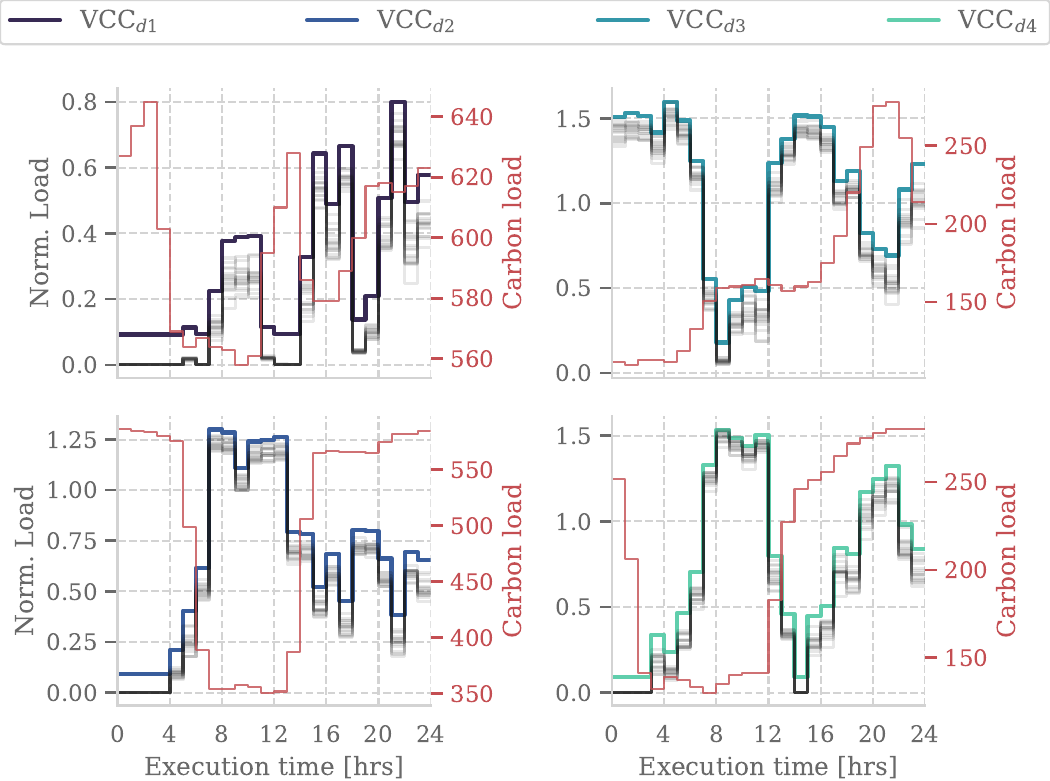}
\caption{Comparing load profiles using real carbon data sourced from\cite{electricitymaps2024carbon} for four distinct Google DC locations.} \label{fig:CarbonCost}
\end{figure}

We use real carbon data for four distinct locations of Google DCs: (i) Berkeley, South Carolina, (ii) Henderson, Nevada, (iii) Saint-Ghislain, Belgium, (iv), Quilicura, Chile. We use the carbon intensity data from Electricity Maps~\cite{electricitymaps2024carbon}, which is the official company that Google sources its carbon data from in real operation. For the presented simulation we used data from the 22nd of February, starting at 10 am. The results are presented in Figure~\ref{fig:CarbonCost}. First of all, we notice that the scheduling strategy exploits the geographic flexibility of compute load to decrease the load in the carbon intense location of Berkeley in South Carolina. Only load that is spatially inflexible is placed here. In addition, we notice that the scheduler also exploits the temporal flexibility of jobs in Henderson, Nevada where carbon intensity greatly drops from hour 8 to 12 am.

\section{Conclusion}\label{sec:Conclusion}
We presented a scheme for spatial and temporal management of flexible computing jobs in large-scale data center fleets which is driven by data, computationally efficient for real-time operation and retains theoretical performance guarantees provided by the DRO solution. The day-ahead planning problem incorporates historical compute load data and future predictions on load and prices to compute a cost-efficient robust aggregate load schedule which is tracked in real-time. Using normalized load profiles from randomly selected Google clusters, we demonstrated that the DRO scheduling strategy outperforms commonly used greedy policies while offering a direct way to trade off performance versus constraint violations through tuning parameters of the CVaR constraint. Future work will investigate a receding-horizon implementation of the DRO policy with real job-level data.

\bibliography{CarbonAwareComputing}

\begin{thebibliography}{10}

\bibitem{masanet2020recalibrating}
E.~Masanet, A.~Shehabi, N.~Lei, S.~Smith, and J.~Koomey, ``Recalibrating global
  data center energy-use estimates,'' {\em Science}, vol.~367, pp.~984--986,
  Feb. 2020.

\bibitem{iea2023data}
{International Energy Agency}, ``Data centres and data transmission networks.''
  [online], July 2023.

\bibitem{gsg2024generational}
{The Goldman Sachs Group}, ``{Generational Growth AI, data centers and the
  coming US power demand surge}.'' [online], 4 2024.

\bibitem{hodgson2024booming}
C.~Hodgson, ``Booming ai demand threatens global electricity supply,'' {\em
  Financial Times}, 2024.
\newblock Accessed: 2024-10-18.

\bibitem{enelx2024how}
{Enel X}, ``How data centers support the power grid with ancillary services,''
  2024.
\newblock Accessed: 2024-10-18.

\bibitem{srg2021microsoft}
{Synergy Research Group}, ``{Microsoft, Amazon and Google Account for Over Half
  of Today’s 600 Hyperscale Data Centers}.''
  \url{https://tinyurl.com/3tz73kv7}, Jan. 2021.
\newblock Accessed: 2024-07-05.

\bibitem{johnson2020carbon}
B.~Johnson, ``Carbon-aware kubernetes: Reducing emissions with smart scaling,''
  Oct. 2020.
\newblock Microsoft Developer Blog.

\bibitem{ramachandran2024announcing}
R.~Ramachandran, ``Announcing the public preview of azure compute fleet.''
  Microsoft, May 2024.
\newblock Accessed: 2024-10-10.

\bibitem{dixit2024retinas}
H.~D. Dixit and J.~Tse, ``Retinas: Real-time infrastructure accounting for
  sustainability.'' Meta Engineering Blog, 2024.
\newblock Accessed: 2024-10-10.

\bibitem{verrus2024how}
Verrus, ``How verrus is powering the data future.'' Verrus News, 2024.
\newblock Accessed: 2024-10-10.

\bibitem{google2024net}
{Google}, ``Net zero carbon: Operating sustainably,'' 2024.
\newblock Accessed: 2024-10-18.

\bibitem{bovera2018economic}
F.~Bovera, M.~Delfanti, and F.~Bellifemine, ``Economic opportunities for demand
  response by data centers within the new italian ancillary service market,''
  in {\em 2018 IEEE International Telecommunications Energy Conference
  (INTELEC)}, vol.~10, pp.~1--8, IEEE, Oct. 2018.

\bibitem{hansson2022potential}
J.~Hansson, ``The potential of data centre participation in ancillary service
  markets in {S}weden,'' Master's thesis, KTH, School of Industrial Engineering
  and Management (ITM), 2022.

\bibitem{wierman2014opportunities}
A.~Wierman, Z.~Liu, I.~Liu, and H.~Mohsenian-Rad, ``Opportunities and
  challenges for data center demand response,'' in {\em International Green
  Computing Conference}, vol.~14, pp.~1--10, IEEE, Nov. 2014.

\bibitem{mehra2024using}
V.~Mehra and R.~Hasegawa, ``Using demand response to reduce data center power
  consumption.'' \url{https://tinyurl.com/msj84hcy}, 2024.
\newblock Accessed: 2024-10-18.

\bibitem{xu2020managing}
M.~Xu and R.~Buyya, ``Managing renewable energy and carbon footprint in
  multi-cloud computing environments,'' {\em Journal of Parallel and
  Distributed Computing}, vol.~135, pp.~191--202, 2020.

\bibitem{abusharkh2017optimal}
M.~Abu~Sharkh, A.~Shami, and A.~Ouda, ``Optimal and suboptimal resource
  allocation techniques in cloud computing data centers,'' {\em Journal of
  Cloud Computing}, vol.~6, Mar. 2017.

\bibitem{dvorkin2024agent}
V.~Dvorkin, ``Agent coordination via contextual regression (agentconcur) for
  data center flexibility,'' {\em {IEEE} Trans. Power Syst.}, pp.~1--11, 2024.

\bibitem{chen2017dglb}
T.~Chen, A.~G. Marques, and G.~B. Giannakis, ``Dglb: Distributed stochastic
  geographical load balancing over cloud networks,'' {\em IEEE Trans. Parallel
  Distrib. Syst.}, vol.~28, no.~7, pp.~1866--1880, 2017.

\bibitem{liu2015greening}
Z.~Liu, M.~Lin, A.~Wierman, S.~Low, and L.~L.~H. Andrew, ``Greening
  geographical load balancing,'' {\em IEEE ACM Transactions on Networking},
  vol.~23, no.~2, pp.~657--671, 2015.

\bibitem{lindberg2022using}
J.~Lindberg, B.~C. Lesieutre, and L.~A. Roald, ``Using geographic load shifting
  to reduce carbon emissions,'' {\em Electric Power Systems Research},
  vol.~212, p.~108586, 2022.

\bibitem{paul2013price}
D.~Paul and W.-D. Zhong, ``Price and renewable aware geographical load
  balancing technique for data centres,'' in {\em 2013 9th International
  Conference on Information, Communications and Signal Processing}, pp.~1--5,
  2013.

\bibitem{breukelman2024carbon}
E.~Breukelman, S.~Hall, G.~Belgioioso, and F.~D{"o}rfler, ``Carbon-aware
  computing in a network of data centers: A hierarchical game-theoretic
  approach,'' in {\em 2024 European Control Conference (ECC)}, pp.~798--803,
  IEEE, 2024.

\bibitem{wang2018optimal}
R.~Wang, Y.~Lu, K.~Zhu, J.~Hao, P.~Wang, and Y.~Cao, ``An optimal task
  placement strategy in geo-distributed data centers involving renewable
  energy,'' {\em IEEE Access}, vol.~6, pp.~61948--61958, 2018.

\bibitem{khosravi2017dynamic}
A.~Khosravi, L.~L.~H. Andrew, and R.~Buyya, ``Dynamic vm placement method for
  minimizing energy and carbon cost in geographically distributed cloud data
  centers,'' {\em IEEE Trans. Sustain. Comput.}, vol.~2, no.~2, pp.~183--196,
  2017.

\bibitem{nadalizadeh2021greenpacker}
Z.~Nadalizadeh and M.~Momtazpour, ``{GreenPacker}: renewable- and
  fragmentation-aware {VM} placement for geographically distributed green data
  centers,'' {\em The Journal of Supercomputing}, vol.~78, pp.~1434--1457, jun
  2021.

\bibitem{goiri2015matching}
{\'I}.~Goiri, M.~E. Haque, K.~Le, R.~Beauchea, T.~D. Nguyen, J.~Guitart,
  J.~Torres, and R.~Bianchini, ``Matching renewable energy supply and demand in
  green datacenters,'' {\em Ad Hoc Networks}, vol.~25, pp.~520--534, 2015.

\bibitem{aksanli2016renewable}
B.~Aksanli, J.~Venkatesh, I.~Monga, and T.~S. Rosing, ``Renewable energy
  prediction for improved utilization and efficiency in datacenters and
  backbone networks,'' in {\em Computational Sustainability}, pp.~47--74,
  Springer, 2016.

\bibitem{khalil2019energy}
M.~I.~K. Khalil, I.~Ahmad, and A.~A. Almazroi, ``Energy efficient indivisible
  workload distribution in geographically distributed data centers,'' {\em IEEE
  Access}, vol.~7, pp.~82672--82680, 2019.

\bibitem{radovanovic2023carbon}
A.~Radovanovi{\'{c}}, R.~Koningstein, I.~Schneider, B.~Chen, A.~Duarte, B.~Roy,
  D.~Xiao, M.~Haridasan, P.~Hung, N.~Care, S.~Talukdar, E.~Mullen, K.~Smith,
  M.~Cottman, and W.~Cirne, ``{Carbon-Aware Computing for Datacenters},'' {\em
  {IEEE} Trans. Power Syst.}, vol.~38, pp.~1270--1280, mar 2023.

\bibitem{TejadaArango2020TPWRS}
D.~A. Tejada{-}Arango, G.~Morales{-}Espa{\~{n}}a, S.~Wogrin, and E.~Centeno,
  ``Power‑based generation expansion planning for flexibility requirements,''
  {\em IEEE Transactions on Power Systems}, vol.~35, no.~3, pp.~2012--2023,
  2020.

\bibitem{Asensio2018TSG}
M.~Asensio, P.~M. de~Quevedo, G.~M. {n}oz{-}Delgado, and J.~Contreras, ``Joint
  distribution network and renewable energy expansion planning considering
  demand response and energy storage—part~ii: Numerical results,'' {\em IEEE
  Transactions on Smart Grid}, vol.~9, no.~2, pp.~667--675, 2018.

\bibitem{VazquezPombo2023TSG}
D.~V. Pombo, J.~Mart{\'{\i}}nez{-}Rico, M.~Carri\'on, and
  M.~Ca{\~{n}}as{-}Carret\'on, ``A computationally efficient formulation for
  flexibility‑enabling generation expansion planning,'' {\em IEEE
  Transactions on Smart Grid}, vol.~14, no.~4, pp.~2723--2733, 2023.

\bibitem{kuhn2019wasserstein}
D.~Kuhn, P.~M. Esfahani, V.~A. Nguyen, and S.~Shafieezadeh-Abadeh,
  ``Wasserstein distributionally robust optimization: Theory and applications
  in machine learning,'' in {\em Operations research \& management science in
  the age of analytics}, pp.~130--166, Informs, 2019.

\bibitem{esfahani2017data}
P.~M. Esfahani and D.~Kuhn, ``{Data-driven distributionally robust optimization
  using the Wasserstein metric: performance guarantees and tractable
  reformulation}s,'' {\em Mathematical Programming}, vol.~171, pp.~115--166,
  jul 2017.

\bibitem{dean2004mapreduce}
J.~Dean and S.~Ghemawat, ``{MapReduce}:simplified data processing on large
  clusters,'' in {\em OSDI'04: Sixth Symposium on Operating System Design and
  Implementation}, (San Francisco, CA), pp.~137--150, 2004.

\bibitem{radovanovic2022power}
A.~Radovanovic, B.~Chen, S.~Talukdar, B.~Roy, A.~Duarte, and M.~Shahbazi,
  ``Power modeling for effective datacenter planning and compute management,''
  {\em IEEE Trans. Smart Grid}, vol.~13, no.~2, pp.~1611--1621, 2022.

\bibitem{google2024efficiency}
{Google}, ``Growing the internet while reducing energy consumption: Water and
  cooling.'' Google Data Centers, 2024.
\newblock Accessed: \today.

\bibitem{gao2014machine}
J.~Gao and R.~Jamidar, ``Machine learning applications for data center
  optimization,'' {\em Google White Paper}, vol.~21, pp.~1--13, 2014.

\bibitem{swissgrid2019transmission}
{Swissgrid Ltd}, ``Transmission code 2019,'' Technical Code 1005/TC, Swissgrid
  Ltd, Aarau, Switzerland, 2019.
\newblock Valid from 07.05.2020.

\bibitem{electricitymaps2024carbon}
{Electricity Maps}, ``{Carbon Intensity Data}.'' [online], 2024.

\bibitem{chen2016robust}
T.~Chen, Y.~Zhang, X.~Wang, and G.~B. Giannakis, ``Robust workload and energy
  management for sustainable data centers,'' {\em IEEE Journal on Selected
  Areas in Communications}, vol.~34, no.~3, pp.~651--664, 2016.

\bibitem{james2019low}
A.~James and D.~Schien, ``A low carbon kubernetes scheduler.,'' in {\em ICT4S},
  2019.

\bibitem{narayanan2017right}
I.~Narayanan, A.~Kansal, and A.~Sivasubramaniam, ``Right-sizing geo-distributed
  data centers for availability and latency,'' in {\em 2017 IEEE 37th
  International Conference on Distributed Computing Systems (ICDCS)},
  pp.~230--240, 2017.

\bibitem{kleywegt2002sample}
A.~J. K. A. S.~T. {Homem-de-Mello}, ``{The Sample Average Approximation Method
  for Stochastic Discrete Optimization},'' {\em {SIAM} Journal on
  Optimization}, vol.~12, pp.~479--502, jan 2002.

\bibitem{subirats2015assessing}
J.~Subirats and J.~Guitart, ``Assessing and forecasting energy efficiency on
  cloud computing platforms,'' {\em Future Generation Computer Systems},
  vol.~45, pp.~70--94, 2015.

\bibitem{villani2009optimal}
C.~Villani {\em et~al.}, {\em Optimal transport: old and new}, vol.~338.
\newblock Springer, 2009.

\bibitem{taleb2010black}
N.~N. Taleb, {\em The black swan : the impact of the highly improbable}.
\newblock New York Times Bestseller, New York: Random House Trade Paperbacks,
  2nd ed., random trade pbk. ed.~ed., 2010.

\bibitem{hota2019data}
A.~R. Hota, A.~Cherukuri, and J.~Lygeros, ``Data-driven chance constrained
  optimization under wasserstein ambiguity sets,'' in {\em 2019 American
  Control Conference (ACC)}, pp.~1501--1506, IEEE, 2019.

\bibitem{tirmazi2020borg}
M.~Tirmazi, A.~Barker, N.~Deng, M.~E. Haque, Z.~G. Qin, S.~Hand,
  M.~Harchol-Balter, and J.~Wilkes, ``Borg: the next generation,'' in {\em
  Proceedings of the fifteenth European conference on computer systems},
  pp.~1--14, 2020.

\bibitem{verma2015large}
A.~Verma, L.~Pedrosa, M.~R. Korupolu, D.~Oppenheimer, E.~Tune, and J.~Wilkes,
  ``Large-scale cluster management at {Google} with {Borg},'' in {\em
  Proceedings of the European Conference on Computer Systems (EuroSys)},
  (Bordeaux, France), 2015.

\bibitem{gurobi}
{Gurobi Optimization, LLC}, ``{Gurobi Optimizer Reference Manual},'' 2023.

\end{thebibliography}

\begin{IEEEbiography}[{\includegraphics[width=1in,height=1.25in,clip,keepaspectratio]{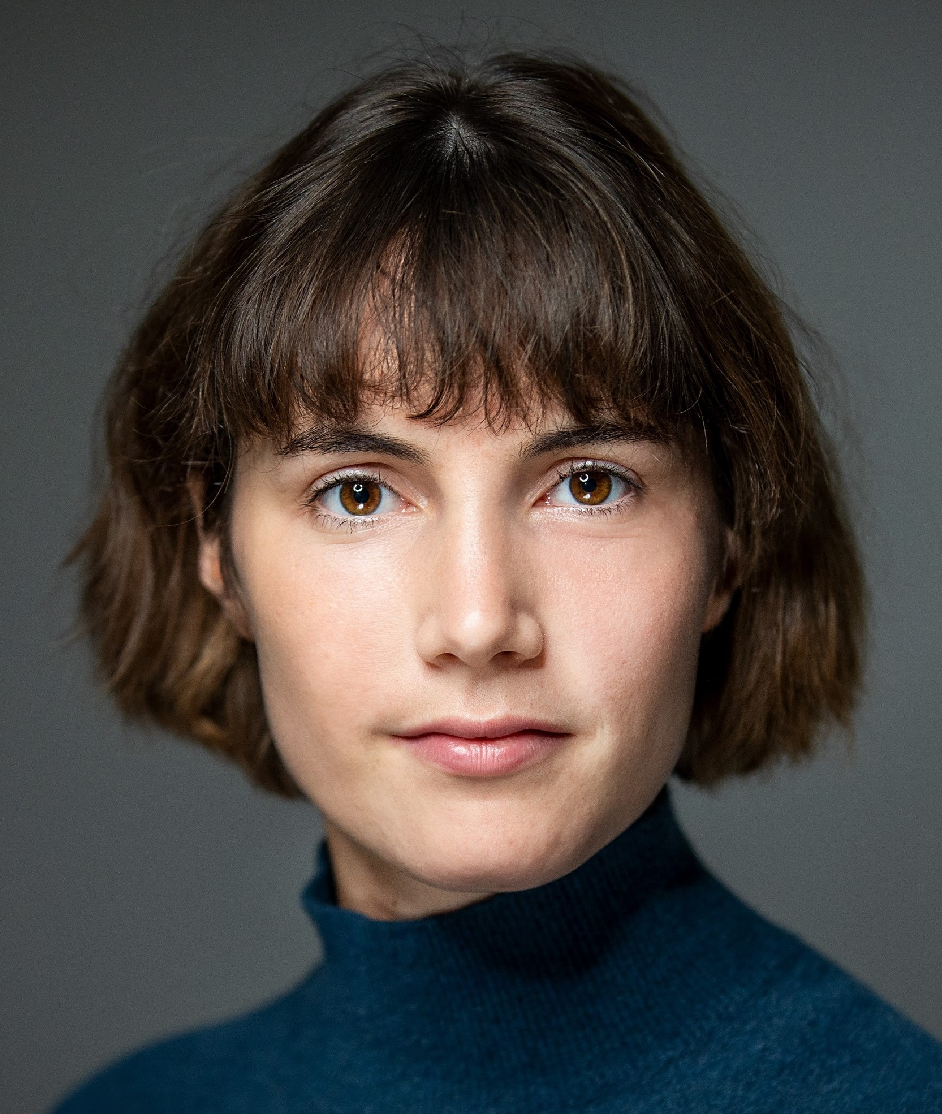}}]{Sophie Hall} is a PhD student at the Automatic Control Laboratory at ETH Zurich. She received the Bachelor's degree in Mechanical Engineering from the University of Surrey, UK, and Nanyang Technological University, Singapore. She completed her Master's degree at ETH Zurich in Biomedical Engineering focusing on modelling and control. She was a finalist for the IFAC NMPC 2024 Young Authors Award. Her research interests revolve around game theory, model predictive control and real-time optimization with applications in network systems such as energy and supply chains.
\end{IEEEbiography}

\begin{IEEEbiography}[{\includegraphics[width=1in,height=1.25in,clip,keepaspectratio]{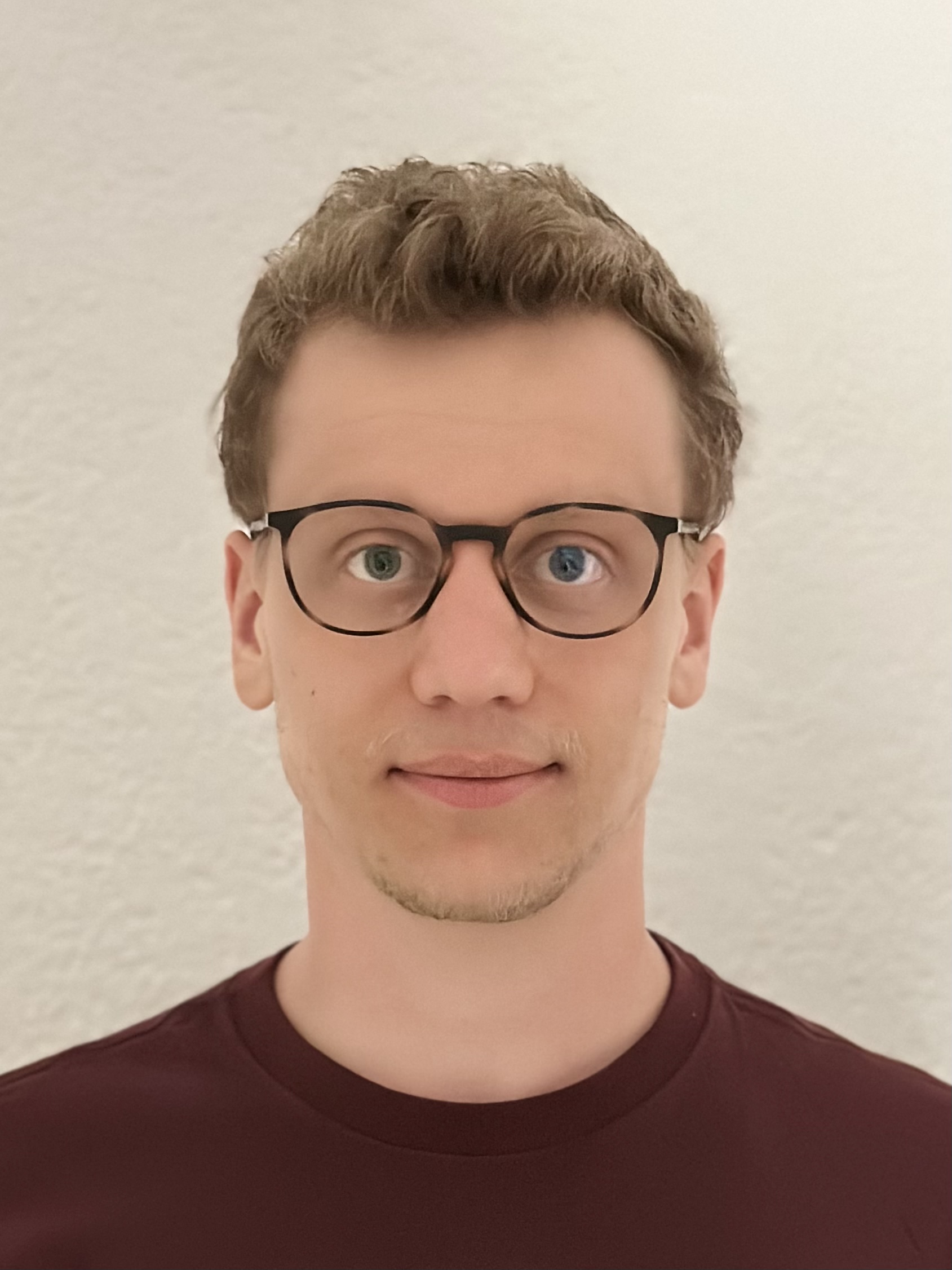}}]{Francesco Micheli} is a PhD student at the Automatic Control Laboratory at ETH Zurich, under the supervision of Prof. J. Lygeros. He received his B.Sc. and M.Sc. degrees in Mechanical Engineering from Politecnico di Milano, Italy, in 2017 and 2018, respectively. His research focuses on safe learning and control, distributionally robust optimization, and robotics.
\end{IEEEbiography}

\begin{IEEEbiography}[{\includegraphics[width=1in,height=1.25in,clip,keepaspectratio]{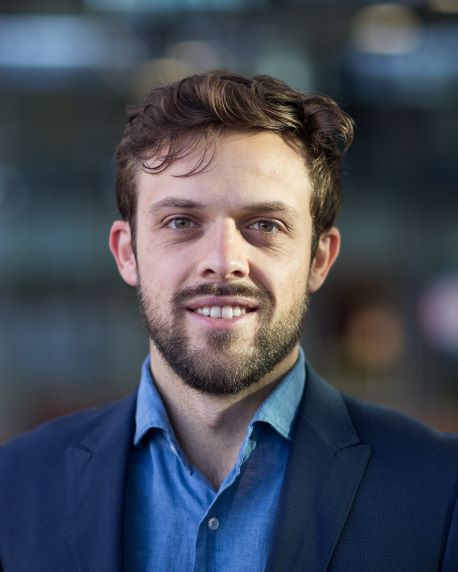}}]{Giuseppe Belgioioso} (Member, IEEE) is an Assistant Professor at the Division of Decision and Control Systems at KTH Royal Institute of Technology, Sweden. He received the Bachelor's degree in Information Engineering in 2012 and the Master's degree (cum laude) in Control Systems in 2015, both at the University of Padova, Italy. In 2020, he obtained the Ph.D. degree in Automatic Control at Eindhoven University of Technology (TU/e), The Netherlands. From 2021 to 2024, he was first a Postdoctoral researcher and then Senior Scientist at the Automatic Control Laboratory, ETH Z\"{u}rich, Switzerland. His research lies at the intersection of optimization, game theory, and automatic control with applications in complex systems, such as power grids and traffic networks.
\end{IEEEbiography}

\begin{IEEEbiography}[{\includegraphics[width=1in,height=1.25in,clip,keepaspectratio]{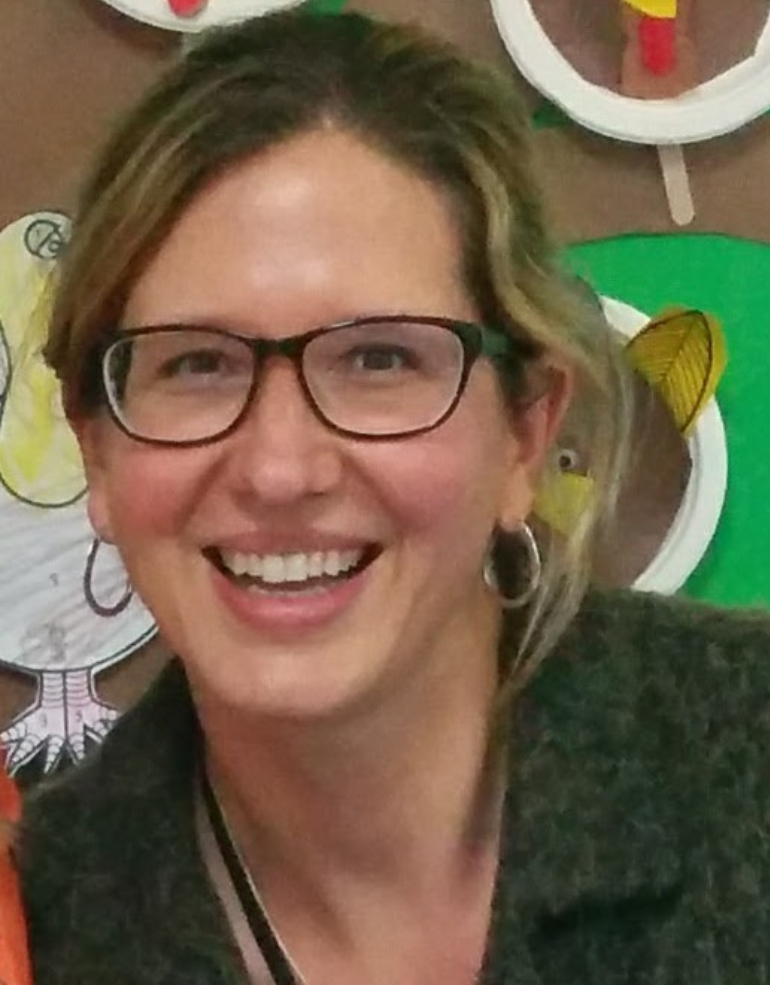}}]{Ana Radovanovi\'{c}} has been a research scientist at Google since early 2008, after she earned her PhD Degree in Electrical Engineering from Columbia University (2005) and worked for 3 years as a Research Staff Member in the Mathematical Sciences Department at IBM TJ Watson Research Center. For more than 10 years, Ana Radovanovi\'{c} has focused all her research efforts at Google on building innovative technologies and business models with two goals in mind: (i) to deliver more reliable, affordable and clean electricity to everyone in the world, and (ii) to help Google become a thought leader in decarbonizing the electricity grid. Nowadays, Ana is widely recognized as a technical lead and research entrepreneur. She is a Senior Staff Research Scientist, serving as a Technical Lead for Energy Analytics and Carbon Aware Computing at Google
\end{IEEEbiography}

\begin{IEEEbiography}[{\includegraphics[width=1in,height=1.25in,clip,keepaspectratio]{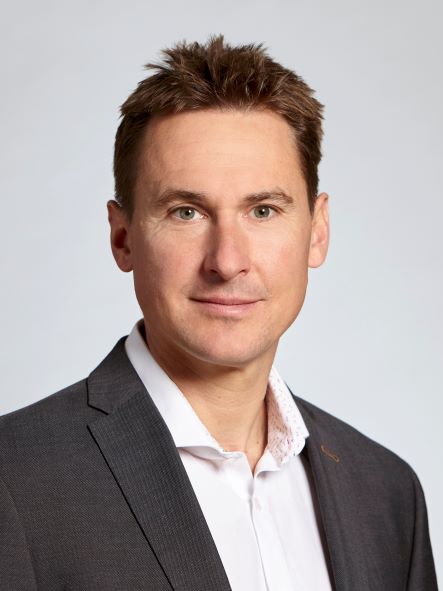}}]{Florian D\"orlfer}
is a Full Professor at the Automatic Control Laboratory at ETH Zurich. He received his Ph.D. degree in Mechanical Engineering from the University of California at Santa Barbara in 2013, and a Diplom degree in Engineering Cybernetics from the University of Stuttgart in 2008. From 2013 to 2014 he was an Assistant Professor at the University of California Los Angeles. He has been serving as the Associate Head of the ETH Zurich Department of Information Technology and Electrical Engineering from 2021 until 2022. His research interests are centered around control, optimization, and system theory with applications in network systems, in particular electric power grids.
\end{IEEEbiography}

\end{document}